\documentclass[twocolumn]{aastex63}

\usepackage{CJKutf8}
\usepackage[scaled]{helvet} 
\usepackage[T1]{fontenc}
\usepackage{hyperref} 
\usepackage{mathtools}
\usepackage{balance}
\usepackage{lineno}
\usepackage{ragged2e}
\justifying

\shorttitle{Vertically resolved magma ocean--protoatmosphere evolution}
\shortauthors{Lichtenberg et al.}
\graphicspath{{./}{figures/}}

\versioninfo{\footnotesize Journal of Geophysical Research: Planets, \textit{Exoplanets: The Nexus of Astronomy and Geoscience}, doi:
\href{https://doi.org/10.1029/2020JE006711}{10.1029/2020JE006711}}

\begin{document}

\title{\large
Vertically resolved magma ocean--protoatmosphere evolution:\\
H$_2$, H$_2$O, CO$_2$, CH$_4$, CO, O$_2$, and N$_2$ as primary absorbers
}

\author[0000-0002-3286-7683]{Tim Lichtenberg}
\affiliation{Atmospheric, Oceanic and Planetary Physics, Department of Physics, University of Oxford, United Kingdom}
\correspondingauthor{Tim Lichtenberg}
\email{tim.lichtenberg@physics.ox.ac.uk}
\author[0000-0002-0673-4860]{Dan J. Bower}
\affiliation{Center for Space and Habitability, University of Bern, Switzerland}
\author[0000-0002-6893-522X]{Mark Hammond}
\affiliation{Department of the Geophysical Sciences, University of Chicago, USA}
\author[0000-0002-5728-5129]{Ryan Boukrouche}
\affiliation{Atmospheric, Oceanic and Planetary Physics, Department of Physics, University of Oxford, United Kingdom}
\author[0000-0003-3968-8482]{Patrick Sanan}
\affiliation{Institute of Geophysics, Department of Earth Sciences, ETH Zurich, Switzerland}
\author[0000-0002-8163-4608]{Shang-Min Tsai}
\affiliation{Atmospheric, Oceanic and Planetary Physics, Department of Physics, University of Oxford, United Kingdom}
\author[0000-0002-5887-1197]{Raymond T. Pierrehumbert}
\affiliation{Atmospheric, Oceanic and Planetary Physics, Department of Physics, University of Oxford, United Kingdom}

\begin{abstract}
The earliest atmospheres of rocky planets originate from extensive volatile release during magma ocean epochs that occur during assembly of the planet. These establish the initial distribution of the major volatile elements between different chemical reservoirs that subsequently evolve via geological cycles.
Current theoretical techniques are limited in exploring the anticipated range of compositional and thermal scenarios of early planetary evolution, even though these are of prime importance to aid astronomical inferences on the environmental context and geological history of extrasolar planets. Here, we present a coupled numerical framework that links an evolutionary, vertically-resolved model of the planetary silicate mantle with a radiative-convective model of the atmosphere. Using this method we investigate the early evolution of idealized Earth-sized rocky planets with end-member, clear-sky atmospheres dominated by either H$_2$, H$_2$O, CO$_2$, CH$_4$, CO, O$_2$, or N$_2$.
We find central metrics of early planetary evolution, such as energy gradient, sequence of mantle solidification, surface pressure, or vertical stratification of the atmosphere, to be intimately controlled by the dominant volatile and outgassing history of the planet. Thermal sequences fall into three general classes with increasing cooling timescale: CO, N$_2$, and O$_2$ with minimal effect, H$_2$O, CO$_2$, and CH$_4$ with intermediate influence, and H$_2$ with several orders of magnitude increase in solidification time and atmosphere vertical stratification.
Our numerical experiments exemplify the capabilities of the presented modeling framework and link the interior and atmospheric evolution of rocky exoplanets with multi-wavelength astronomical observations.
\begin{center}
\hspace{1.3cm}PLAIN LANGUAGE SUMMARY
\end{center}
The climate and surface conditions of rocky planets are sensitive to the composition of their atmospheres, but the origins of these atmospheres remain unknown. During the final stages of planetary accretion, when the whole of the planet solidifies from lava to solid rock, volatiles can rapidly cycle between interior and atmosphere, and control the cooling properties of the planet. In order to understand how cooling planets with different volatile compositions evolve, we use computer simulations to analyze their thermal evolution, outgassing, and observable signatures. We find the planetary cooling sequence to differ by orders of magnitude depending on the primary volatile. CO, N$_2$, and O$_2$ rapidly outgas, but allow planets to solidify efficiently. H$_2$O, CO$_2$, and CH$_4$ build an intermediate class with substantially delayed cooling, but varying outgassing rate. H$_2$ most efficiently inhibits solidification relative to the other cases. All considered volatiles show distinctive evolution of the atmosphere and interior and display different atmospheric signals that can be measured by astronomical surveys. Deviations in the geological properties of the planetary interior similarly manifest in the atmospheric signal. Future observations may isolate distinctive features in exoplanet atmospheres to infer interior state and composition.
\vspace{0.99cm}
\end{abstract}

\section{Introduction} \label{sec:intro}

The debate surrounding the formation and long-term evolution of rocky planets has been dominated by the wealth of data obtained from the terrestrial planets and planetary materials in the Solar System. However, current and upcoming astronomical surveys of planet-forming circumstellar disks and evolved extrasolar planets will increase the available data greatly. For instance, astrochemical studies of ever-higher precision deliver clues to the origins of major atmophile elements during the main planet formation era \citep{2020arXiv200612522V,2020ApJ...898...97B}. Ongoing \citep{2015JATIS...1a4003R,2020MNRAS.493..536D} and future \citep{HABEX_StudyReport2019,ORIGINS_StudyReport2019,LUVOIR_StudyReport2019,2019AJ....158...83A,2019arXiv190801316Q} exoplanet surveys will reveal statistically meaningful characteristics of Earth-sized planets on temperate orbits. Having so far only glimpsed the richness and diversity of the exoplanet census, statistical evaluation of basic physical properties---such as mass, radius, and atmospheric features---teach us that individual rocky planets differ substantially from one another and from the terrestrial planets of the Solar System \citep{2019AREPS..47..141J,2019AnRFM..51..275P,2019AREPS..47...67O}. Shaped by extreme post-accretionary processes, the impinging radiation of close-in planets can strip away their volatile envelopes, leaving only dessicated, bare rocks \citep{2013ApJ...775..105O,2014ApJ...792....1L,2014ApJ...795...65J,2017AJ....154..109F}, with permanent day- and nightsides \citep{2019Natur.573...87K} or in eternal runaway greenhouse states \citep{2019AA...628A..12T}. Further out, being born from volatile-rich building blocks, solid-dominated planets may resemble scaled-up versions of the icy moons of the outer Solar System \citep{2003ApJ...596L.105K,2004Icar..169..499L}, where high-pressure ice phases at the mantle-atmosphere interface inhibit Earth-like geochemical cycling of nutrients \citep{2017SSRv..212..877N,2018ApJ...864...75K,2020SSRv..216....7J}. For the more massive super-Earths, the crushing pressures at depth generate a supercritical fluid of equilibrated vapor and rock \citep{2019ApJ...887L..33K,2020ApJ...891L...7M}, rendering a surface absent in the traditional sense. 

These extreme conditions and processes produce compositional differences between exoplanets that dwarf those between the Solar System terrestrial planets, where from an exoplanet viewpoint the compositional phase space of the Solar System planets is approximately solar \citep{1995ChGeo.120..223M,2009ARAA..47..481A} except for the volatile budget \citep{zahnle2020creation}. The delivery and redistribution of volatile compounds during planetary evolution constitutes a major gap in our knowledge, yet is crucial for the origin and evolution of terrestrial atmospheres. The uncertainty in volatile behavior limits both our understanding of the planetary environmental context of early Earth and the evolutionary history of rocky exoplanets. Relative to larger, gas or ice-phase dominated worlds, the surface and climatic conditions of rocky planets are sensitive to a variety of physical and chemical processes that operate during their formation, on their precursors bodies, and after their formation. Planetary embryos are anticipated to undergo orbital migration during the disk phase \citep{2019AA...624A.109B} and iceline positions evolve \citep{2000ApJ...528..995S,2020MNRAS.495.3160O} during episodic planetesimal formation \citep{2018AA...614A..62D}, and thus forming planets can sample an evolving mixture of volatiles in various physical and chemical states during accretion \citep{2011ApJ...743L..16O,2018AA...613A..14E}. Recent work suggests that forming protoplanets and their precursors experience a significant degree of high-energy processing \citep{2019GeCoA.260..204S,Fegley2020a,2020Icar..34713772B} during planetary formation. Planetesimals and protoplanets evolve due to impacts \citep{2016ApJ...821..126Q,2020arXiv200704321K,2020MNRAS.496.1166D} and internal radiogenic heating \citep{2016Icar..274..350L,2019NatAs...3..307L,2019EPSL.507..154L}, both of which dramatically alter the thermal budget and volatile content of young rocky worlds. The composition of early- and late-accreted material can alter the initial oxidation state, and thus chemical speciation of the upper mantle and atmosphere \citep{2014EPSL.403..307G,2020NatSR..1010907O,2020PSJ.....1...11Z}, which is relevant for the planetary environment of prebiotic chemistry \citep{Benner2019,2019Icar..329..124R,2020SciA....6.3419S}.

The climatic conditions of young, rocky worlds are therefore established by the extreme cycling of volatiles between planetary reservoirs when the majority of the planet, including its iron core and silicate mantle, is molten. Liquid--gas interactions facilitate rapid chemical equilibration and volatile exchange. The magma ocean phase is thus a crucial evolutionary link between the present state of exoplanets and their formation history \citep{2009ApJ...704..770M,2015ApJ...806..216H,2014ApJ...784...27L,2016ApJ...829...63S,2019AA...621A.125B}. The abundance of water-rich phases in carbonaceous chondritic meteorites and the importance of H$_2$O as a greenhouse gas motivated pioneering work on steam atmospheres above a magma ocean \citep{1986Natur.322..526M,1986Natur.319..303M,1988Icar...74...62Z}, which has since been extended to the early thermal and compositional evolution of young protoplanets \citep{elkins2008linked,2011ApSS.332..359E,2012AREPS..40..113E,2013Natur.497..607H,2015ApJ...806..216H,lebrun2013thermal,2017JGRE..122.1458S,2017JGRE..122.1539M,2016ApJ...829...63S,2016SSRv..205..153M,2017RSPTA.37550394T,2018AARv..26....2L,2019ApJ...875...31K,2019ApJ...875...11N,2019AA...631A.103B,Barth2020,KiteBarnett2020PNAS}. However, meteoritic materials are debris of the specific pathway of the planet formation process in the Solar System \citep{2016EPSL.435..136W,2018Icar..302...27L} and may differ substantially from the compositional variety in exoplanetary systems. In addition, the compositional deviations between Solar System chondritic groups result in complex transitions in outgassing behavior \citep{2007Icar..186..462S,2010Icar..208..438S}, such that the outgassing speciation of chondrite mixtures cannot be deduced from simple superpositions of individual chondrite outgassing speciation \citep{2017ApJ...843..120S}. Furthermore, the chemical speciation of protoatmospheres is a result of the dynamic and chemical evolution of the silicate mantle during planet solidification, which acts as a reservoir into which major greenhouse gases partition to different degrees during the magma ocean phase. This sequestering into the interior decouples the evolving climate state of the planet from the volatile speciation of its building blocks.

Evidently, we require advanced tools to assess the planetary state shortly after planetary formation to provide the starting point for the long-term climatic and geodynamic evolution, and allow us to explore the range of planetary environments that may form. Aiming to further reduce Solar System-centric assumptions, we here present the evolution and solidification of protoplanets under diverse climatic settings. We focus on major greenhouse gases and compounds that have been suggested to transiently dominate the chemical composition of protoatmospheres. We explore single-species atmospheres composed of H$_2$, H$_2$O, CO$_2$, CH$_4$, CO, O$_2$, and N$_2$, since these volatiles can become the dominant absorbers under plausible scenarios of planetary accretion and early evolution.
For instance, H$_2$ can affect the long-term climate due to collision-induced absorption \citep{2011ApJ...734L..13P,2019ApJ...881..120K} and its interaction with N$_2$ has been proposed as a contributor to greenhouse warming on early Earth \citep{2013Sci...339...64W}. Furthermore, H$_2$ is abundant during planet formation and can create transient reducing climates on young planets. Even though H$_2$O~is a trace species in the present-day atmosphere of Earth, it has a strong greenhouse effect which has been routinely discussed in the context of magma ocean protoatmospheres \citep{1986Natur.319..303M,1986JGR....91E.291A,1988Icar...74...62Z}. The often-remarked influence of H$_2$O~ on the tectonic regime of Earth- and super-Earth-sized planets suggest that the near-surface geochemistry is sensitive to its initial abundance \citep{2017RSPTA.37550394T,2018RSPTA.37680109S}

CO$_2$ is another major greenhouse species typically included in models of coupled magma ocean and atmosphere evolution \citep{2017JGRE..122.1458S,2019AA...631A.103B}. However, the major phase and partitioning behavior of carbon during planet formation and magma ocean cooling are poorly constrained \citep{2012EPSL.341...48H,dalou2017nitrogen,Fischer2020} despite their relevance for the climate of rocky planets. For instance, CO$_2$ can influence the rate of planetary water loss by affecting the cold-trap temperature and the rate of hydrogen escape \citep{2013ApJ...778..154W}, and may be the major factor that determines the past and present climates of Earth or Mars analogues \citep{1997Sci...278.1273F,2017GeoRL..44..665W}. The dominant species in Earth's present-day atmosphere is N$_2$, and but its major biogeochemical cycles remain debated \citep{2016EPSL.447..103W,2018AsBio..18..897L,2018AARv..26....2L}. Specific planetary settings may render CH$_4$, CO, and O$_2$ the major species in planetary atmospheres for transient epochs. For instance, H$_2$-dominated atmospheres with a mole fraction of H/C of 0.5 may cool down to a CH$_4$-dominated atmosphere at reduced conditions \citep{2012EPSL.341...48H}. CO-dominated atmospheres have been suggested as the result of CO capture from extrasolar debris disks \citep{2020NatAs.tmp...66K}. Finally, O$_2$ atmospheres may result from water photolysis and subsequent H$_2$ loss in steam atmospheres \citep{2015AsBio..15..119L,2016ApJ...829...63S,2018AJ....155..195W}.

Compositional influences on the radiative balance of protoatmospheres will need to be taken into account to probe and refine the potential orbital transition from Venus-like, primordial runaway greenhouse states, which cool via water dissociation and loss to space, to planets that can cool in an Earth-like fashion by radiating away primordial heat \citep{2013Natur.497..607H}. Long-lived magma ocean states may be directly observable with near-future astronomical instrumentation \citep{2014ApJ...784...27L,2015ApJ...806..216H,2019AA...621A.125B} and link the present-day climates of extrasolar rocky planets with the ancient climate of Hadean Earth. This would allow crucial insights on the thermal and compositional state of young exoplanets and further help constrain the chronology of terrestrial planets in the Solar System \citep{Fegley2020a}. Connecting the thermo-compositional chronology and histories of exoplanets with observations \citep{2016ApJ...829...63S,2018AJ....155..195W} is an essential step to bridge astronomy and planetary science. The goal of this work is thus to demonstrate how a sophisticated coupled model of interior and atmosphere interaction can deliver new understanding in how early climate states are derived from the magma ocean epoch of terrestrial planet evolution. We simulate the energetic and compositional evolution of rocky planets from a fully-molten state to their transition to long-term geodynamic and climatic evolution. We describe our coupled numerical framework in Sect. \ref{sec:methods} and present results in Sect. \ref{sec:results}. We discuss the implications of the work presented here, the main limitations of our approach, and prospects for further studies in Sect. \ref{sec:discussion}, and conclude in Sect. \ref{sec:conclusions}.

\section{Methods} \label{sec:methods}

\begin{figure*}[htb]
\centering
\includegraphics[width=0.99\textwidth]{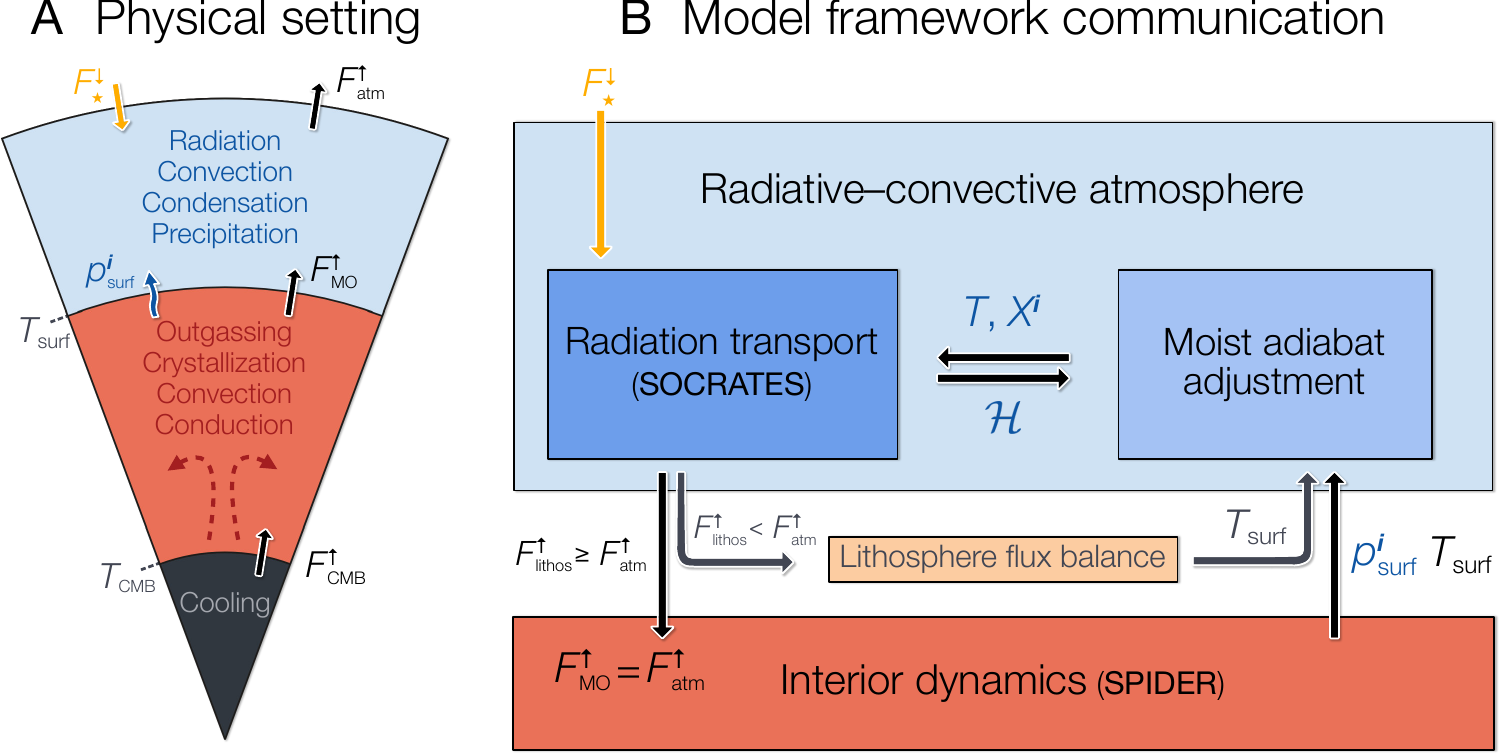}
\caption{\textsf{Schematic of the coupled interior--atmosphere model framework. (\textbf{A}) Magma ocean domain with dominant processes and nominal direction of energy fluxes. (\textbf{B}) Computational model internal communication. Arrow directions show information exchange. Boundary conditions for the interior sub-module are the temperature at the core--mantle boundary $T_{\mathrm{\scriptscriptstyle CMB}}$, and the net heat loss to space $F^{\uparrow}_{\mathrm{MO}} = F^{\uparrow}_{\mathrm{atm}}$, which is computed by the atmosphere sub-module. The interior code computes the boundary conditions for the atmosphere: the surface temperature $T_{\mathrm{surf}}$, and the surface partial pressure from volatile outgassing $p^{i}_{\mathrm{surf}}$. During late-stage cooling, the net heat loss of the planet is controlled by the heat flux in the near-surface (conductive) lithosphere ($F^{\uparrow}_{\mathrm{lithos}}$). In this case, $T_{\mathrm{surf}}$ is computed by equilibrating lithospheric and atmospheric heat fluxes. The atmospheric profile is constructed iteratively according to the adiabatic lapse rate and the atmospheric heating rate from radiative transfer. The top-of-atmosphere boundary is the stellar flux at present time and orbit of the planet.}}
\label{fig:framework}
\end{figure*}

In order to simulate the thermal evolution of a magma ocean and its outgassing atmosphere, we couple the interior dynamics code \textsc{spider} \cite[\textit{Simulating Planetary Interior Dynamics with Extreme Rheology}]{2018PEPI..274...49B} to the atmosphere radiative transfer code \textsc{socrates} \cite[\textit{Suite of Community Radiative Transfer codes based on Edwards and Slingo}]{edwards1996socrates}. Both codes are stand-alone and are executed and mediated in a serial loop by a \texttt{python}-based framework, which synchronizes initial and boundary conditions between the two codes during runtime (Fig. \ref{fig:framework}). The major information exchanged between the interior and the atmosphere is the heat flux and surface temperature at the mantle--atmosphere interface, and the distribution of volatile species between the two reservoirs. We subsequently introduce the individual model components and elucidate how information is communicated between them. 

\subsection{Coupled atmosphere-interior model}
\label{sec:coupled_model}
\subsubsection{Interior}
During planetary solidification the rocky planetary mantle undergoes a dynamic transition from a fully molten and turbulent magma ocean to solid-state convection dominated by viscous creep. In order to capture this evolution, \textsc{spider} \citep{2018PEPI..274...49B,2019AA...631A.103B} follows the interior evolution of an initially fully molten planet during cooling and crystallization by solving the energy conservation equation for a spherically symmetric (1-D) mantle that is discretized using the finite volume method: 
\begin{linenomath*}\begin{equation}
\int_{V} \rho T \frac{\partial S}{\partial t} d V=-\int_{A} F^{\uparrow}_{\mathrm{MO}} \cdot d A+\int_{V} \rho H d V,
\end{equation}\end{linenomath*}
with specific entropy $S$, density $\rho$, temperature $T$, directional heat flux $F^{\uparrow}_{\mathrm{MO}}$, surface area $A$, internal heat generation per unit mass $H$, time $t$, and volume $V$. The heat flux is determined by four dominant energy transport mechanisms that operate in a semi-molten mantle:
\begin{linenomath*}\begin{equation}
F^{\uparrow}_{\mathrm{MO}}=F^{\uparrow}_{\mathrm{conv}}+F^{\uparrow}_{\mathrm{mix}}+F^{\uparrow}_{\mathrm{cond}}+F^{\uparrow}_{\mathrm{grav}}, \label{eq:fluxes}
\end{equation}\end{linenomath*}
with the directional sum of the individual heat flux contributions from convection $F^{\uparrow}_{\mathrm{conv}}$, latent heat transport by mixing of melt and solid $F^{\uparrow}_{\mathrm{mix}}$, conduction $F^{\uparrow}_{\mathrm{cond}}$, and gravitational separation of melt and solid by permeable flow $F^{\uparrow}_{\mathrm{grav}}$. Note that the individual and total heat flux contributions in Eq. \ref{eq:fluxes} can be negative, and thus the model can chart the evolution of both cooling planets ($F^{\uparrow}_{\mathrm{MO}} > 0$) and planets that are heated from the top ($F^{\uparrow}_{\mathrm{MO}} < 0$) such as the substellar hemisphere of tidally-locked super-Earths that lack an atmosphere. Convection is modeled using mixing length theory by determining a local eddy diffusivity following \citet{abe93,abe95,abe97}. Compared to 0-D approaches based on boundary layer theory, mixing length formulations allow for locally-determined variations in temperature and composition at all depths in a planetary mantle. The convective heat flux is
\begin{linenomath*}\begin{equation}
F^{\uparrow}_{\mathrm{conv}} = -\rho T \kappa_{h} \frac{\partial S}{\partial r},
\end{equation}\end{linenomath*}
with eddy diffusivity $\kappa_\mathrm{h} \sim u \mathcal{L}_{\mathrm{m}}$, characteristic convective velocity $u$, and mixing length $\mathcal{L}_{\mathrm{m}}$. Our nominal model uses a constant mixing length, but we also compute cases where the mixing length is equal to the distance to the nearest mantle interface (i.e., core-mantle boundary or planetary surface) \citep{SC97}. Energy is also transported due to the displacement of melt and solid by convection, which causes the release and absorption of latent heat:
\begin{linenomath*}\begin{equation}
F^{\uparrow}_{\mathrm{mix}} = -\rho T \Delta S_{\mathrm{fus}} \kappa_{h} \frac{\partial \phi}{\partial r},
\end{equation}\end{linenomath*}
with entropy of fusion $\Delta S_{\mathrm{fus}} = S_{\mathrm{liq}} - S_{\mathrm{sol}}$, liquidus $S_{\mathrm{liq}}$, solidus $S_{\mathrm{sol}}$, and silicate melt fraction $\phi$. Conduction is
\begin{linenomath*}\begin{equation}
F^{\uparrow}_{\mathrm{cond}} = -\rho T \kappa_t\left(\frac{\partial S}{\partial r}\right)-\rho c \kappa_t\left(\frac{\partial T}{\partial r}\right)_{S},
\end{equation}\end{linenomath*}
with thermal diffusivity $\kappa_t$ and adiabatic temperature gradient $(\partial T/\partial r)_\mathrm{S}$. Finally, molten rock can separate from its surrounding solid rock matrix by permeable flow, and crystals in liquid-dominated domains can settle or float depending on their density contrast. This exchange of energy is captured by the gravitational separation flux:
\begin{linenomath*}\begin{equation}
F^{\uparrow}_{\mathrm{grav}} = a_\mathrm{g}^{2} g \rho\left(\rho_{\mathrm{liq}}-\rho_{\mathrm{sol}}\right) \zeta_{\mathrm{grav}}(\phi) T \Delta S_{\mathrm{fus}} / \eta_{m},
\end{equation}\end{linenomath*}
with grain size $a_\mathrm{g}$, melt viscosity $\eta_\mathrm{m}$, density of melt at the liquidus ($\rho_{\rm liq}$), density of solid at the solidus ($\rho_{\rm sol}$), and flow mechanism factor $\zeta_{\mathrm{grav}}(\phi)$. $\zeta_{\mathrm{grav}}(\phi)$ parameterizes the flow law as a function of melt fraction, to represent the different regimes from crystal-bearing suspensions at high melt fraction to porous flow at low melt fraction \citep[see][]{2018PEPI..274...49B}. 

We assume the silicate mantle to be isochemical, represented by MgSiO$_3$. The thermophysical properties of the pure melt and solid phase are parameterized according to \citet{2018PEPI..278...59W} and \citet{2009JGRB..114.1203M}, respectively. To derive the material properties for the melt-solid aggregate we assume volume additivity when computing the two-phase adiabat. Experiments on synthetic chondritic composition provide the melting curves in the lower mantle \citep{ABL11} and are smoothly joined to the upper mantle melting curves from \citet{2013Natur.497..607H}, which are based on compiled experimental data on melting of dry peridotite KLB-1 \citep{1986JGR....91.9367T,1994JGR....9917729Z,1996JGR...101.8271H,1998Sci...281..243Z,2002E&PSL.197..117T}. The melting curves are assumed to be invariant, hence we neglect the potential depression of the solidus by enrichment of volatiles in the residual melt during solidification.

The geophysical (iron) core is not explicitly modelled, but rather a boundary condition is derived at the core-mantle boundary based on energy balance:
\begin{linenomath*}\begin{equation}
\frac{\mathrm{d} T_{\mathrm{\scriptscriptstyle CMB}}}{\mathrm{d} t}=-\frac{4 \pi r_{\mathrm{\scriptscriptstyle CMB}}^{2} F_{\mathrm{\scriptscriptstyle CMB}}}{m_{\mathrm{core}} c_{\mathrm{core}} \widehat{T}_{\mathrm{core}}},
\end{equation}\end{linenomath*}
with core mass $m_\mathrm{core}$, core heat capacity $c_\mathrm{core}$, thermal structure correction factor $\widehat{T}_\mathrm{core} = T_\mathrm{core}/T_\mathrm{\scriptscriptstyle CMB}$, mass-weighted effective temperature of the core $T_\mathrm{core}$, temperature $T_\mathrm{\scriptscriptstyle CMB}$, radius $r_\mathrm{\scriptscriptstyle CMB}$, and heat flux $F_\mathrm{\scriptscriptstyle CMB}$ at the core-mantle boundary (CMB). The boundary condition states that the net thermal energy of the core decreases as a consequence of heat removal from the CMB by the silicate magma ocean. This condition does not account for crystallisation of the inner core or heat sources within the core \citep{2017GeCoA.199....1C}. The core is assumed to be isentropic (vigorously convecting), which gives rise to a thermal structure correction factor $\hat{T}_\mathrm{core}=1.147$. Hence for our initial mantle thermal profile, the initial core temperature $T_\mathrm{core}=5400$ K.
\subsubsection{Interior--atmosphere interface}
The cooling rate of a mostly molten magma ocean is limited by the ability of the atmosphere to radiate heat to space. Hence during this stage the flux from the interior (upper boundary condition) is $F^{\uparrow}_{\mathrm{MO}} \equiv F^{\uparrow}_{\mathrm{atm}}$, where $F^{\uparrow}_{\mathrm{atm}}$ is computed by the atmosphere sub-module. The surface temperature $T_\mathrm{surf}$ decreases and the mantle cools, driving a redistribution of volatile species $i$ between the interior and atmosphere through outgassing. Hence a new surface temperature and distribution of volatiles in the atmosphere is derived, which are then used to update the thermal boundary condition and atmospheric composition of the atmosphere sub-module. During the later stages of crystallization, a stiff viscous lid forms once the surface temperature drops below the rheological lockup point such that the interior heat flux can decrease below values that can theoretically be sustained by the atmosphere. Cooling of the interior to space is then inhibited due to the comparatively slow rate of energy transfer through the conductive lid (lithosphere), rather than blanketing by the atmosphere. To account for this we decouple the temperature profile of the lithosphere from the internal profile that is spatially resolved by \textsc{spider}. The surface temperature is then updated by solving for conductive transport in a shallow surface region,
\begin{linenomath*}\begin{equation}
\mu_\mathrm{l} \frac{\mathrm{d}T_\mathrm{surf}}{\mathrm{d}t} = F^{\uparrow}_\mathrm{atm} - F^{\uparrow}_\mathrm{lithos} \leq \epsilon,
\end{equation}\end{linenomath*}
with the heat capacity of the near-surface layer $\mu_\mathrm{l}$, the lithosphere heat flux $F^{\uparrow}_\mathrm{lithos}$ computed from the energy gradient near the surface using \textsc{spider}, and convergence factor $\epsilon = 1$ W m$^{-2}$. The atmosphere is then evolved until it equilibrates with the interior heat flux (Fig.~\ref{fig:framework}). When a boundary skin is considered (Fig.~\ref{fig:interior_params}, iv), we account for the ultra-thin thermal boundary layer at the surface that offsets the surface temperature from the mantle potential temperature by several hundred Kelvin. The relationship between the boundary layer temperature drop and the surface temperature is derived by equating the heat flux in an ultra-thin (numerically unresolved) viscous boundary layer with radiation at the surface. For surface temperatures between 1400 K and 2000 K, we then \emph{a priori} solve for the thermal boundary layer structure at steady-state using the heat flux balance and representative material properties of the melt. This determines that the boundary layer temperature drop scales with the surface temperature cubed with a constant of proportionality of $10^{-7}$ K$^{-2}$ \cite[Eq.~18]{2018PEPI..274...49B}.
\begin{table}[tb]
\centering
\begin{tabular}{llrrl}
Molecule  & $\alpha$ (ppmw/Pa)    & $\beta$ (non-dim.) \\ \hline
H$_2$O    & $1.033 \times 10^{0}$   & $1.747$ \\
H$_2$    & $2.572 \times 10^{-6}$  & $1.000$ \\
CO$_2$    & $1.937 \times 10^{-9}$  & $0.714$ \\
CH$_4$   & $9.937 \times 10^{-8}$  & $1.000$ \\
N$_2^-$   & $7.416 \times 10^{+1}$  & $4.582$ \\
N$_2$   & $7.000 \times 10^{-5}$  & $1.800$ \\
CO     & $1.600 \times 10^{-7}$  & $1.000$ \\
O$_2$    & --  & -- \\
\end{tabular}
\caption{\textsf{
Partitioning data and Henry coefficient fit parameters as shown in Fig. \ref{fig:solubilities}. Data references --
H$_2$O: \citet{silver1990influence,holtz1995h2o,moore1998hydrous,yamashita1999experimental,gardner1999experimental,liu2005solubility};
H$_2$: \citet{gaillard2003rate,hirschmann2012solubility};
CO$_2$: \citet{mysen1975solubility,stolper1988experimental,pan1991pressure,blank1993solubilities,dixon1995experimental};
CH$_4$: \citet{ardia2013solubility,keppler2019graphite};
N$_2$: \citet{libourel2003nitrogen,li2013nitrogen,dalou2017nitrogen,mosenfelder2019nitrogen};
CO: fit constants from \citet{yoshioka2019carbon};
O$_2$: see main text. 
N$_2^-$ indicates the partition coefficients for nitrogen species under reduced conditions with a redox state below the iron-w\"ustite buffer (IW), $f$O$_2 \lesssim$ IW.
}}
\label{tab:2}
\end{table}

\begin{figure}[htb]
\centering
\includegraphics[width=0.49\textwidth]{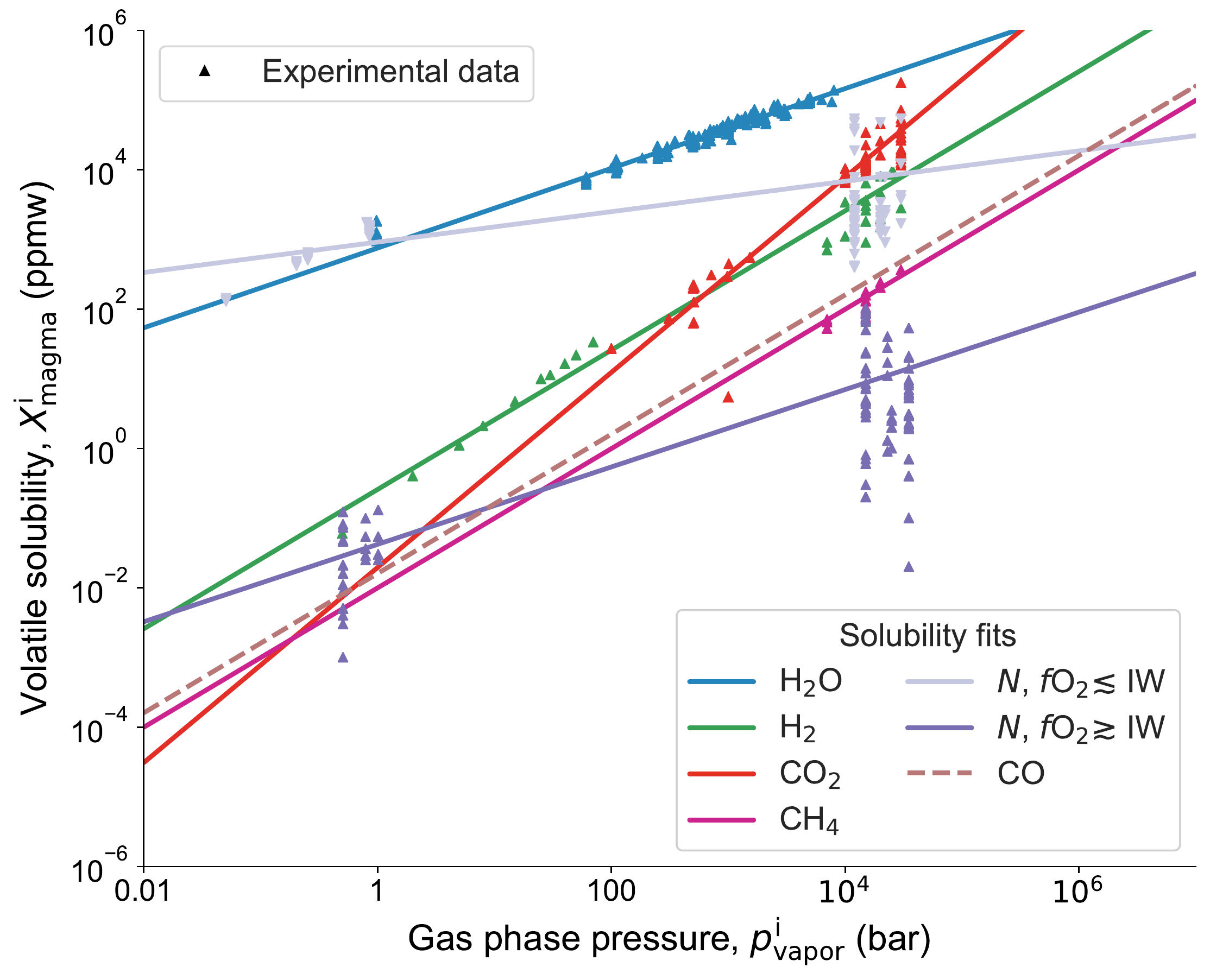}
\caption{\textsf{
Relationship between ambient pressure and solubility in silicate melts for the considered volatiles from petrological experiments. Data references and fit coefficients $\alpha_i$ and $\beta_i$ are given in Tab. \ref{tab:2}. Nitrogen solubility is independently calculated where the oxygen fugacity ($f$O$_2$) of the magma is below or above the iron-w\"ustite buffer (IW), respectively.
}}
\label{fig:solubilities}
\end{figure}

A solubility law relates the abundance of a volatile $i$ in the interior (in melt) to its abundance in the atmosphere, according to the partial pressure of the volatile at the surface: 
\begin{linenomath*}\begin{equation}
p_{\mathrm{surf}}^i=\left(\frac{X^{i}_{\mathrm{magma}}}{\alpha_i}\right)^{\beta_i},
\end{equation}\end{linenomath*}
with the surface partial pressure $p_{\mathrm{surf}}^i$, abundance in the melt $X^{i}_{\mathrm{magma}}$, and Henrian fit coefficients $\alpha_i$ and $\beta_i$ of species $i$ from petrologic data (Tab. \ref{tab:2}, Fig. \ref{fig:solubilities}). In this compilation we introduce two different relations for nitrogen to illustrate its strongly diverging solubility behaviour above and below the iron-w\"ustite buffer \citep{libourel2003nitrogen}. The chemical composition (as quantified by its redox state, $f$O$_2$) of the magma ocean controls the speciation of outgassed volatiles \citep[e.g.]{2018RvMG...84..393S}. These chosen relations hence serve to illustrate the solubility behaviour for a mantle composition that allows the stability of given volatile species.

The partial pressure of a volatile $p_{\rm surf}^i$ then relates to the mass of the volatile in the atmosphere \citep{2019AA...631A.103B}:
\begin{linenomath*}\begin{equation}
m_{i}^{g}=4 \pi R_{p}^{2}\left(\frac{\mu_{i}}{\overline{\mu}_{\mathrm{atm}}}\right) \frac{p_{\mathrm{surf}}^i}{g},
\end{equation}\end{linenomath*}
with planetary radius at the interior--atmosphere interface $R_{p}$, molar mass $\mu_{i}$ of species $i$, mean atmospheric molar mass $\overline{\mu}_{\mathrm{atm}}$, and gravity $g$. We assume O$_2$ to be insoluble and thus its atmospheric mass is constant with time. Our treatment of degassing assumes the respective volatile to be in solution equilibrium between the magma ocean and the overlying atmosphere. If outgassing from the magma ocean would be governed by bubble nucleation instead, the degassing rate may decrease \citep[cf.]{2013Natur.497..607H,2018SSRv..214...76I}.

\subsubsection{Atmosphere}
The atmosphere sub-module of the coupled framework establishes the atmospheric thermal and compositional profile, based on the surface partial pressures $p^\mathrm{i}_\mathrm{surf}$ of the volatile species $i$ and surface temperature $T_\mathrm{surf}$. In the earliest, hottest phase, the atmospheric profile follows a dry adiabat, but with cooling the upper atmosphere saturates, condenses, and follows the saturation vapor curve \citep{PierrehumbertBook2010}. Because the net energy flux to space depends on the radiating layer within the atmosphere, it is important to consider the influence of the change in atmospheric lapse rate when temperatures are cool enough for condensation to occur. We use a fourth-order Runge-Kutta integrator to compute the atmosphere lapse rate by integrating upwards from the planetary surface where $T=T_\mathrm{surf}$:
\begin{linenomath*}\begin{eqnarray}
\frac{d \ln T}{d \ln P} &= R/c_{p,i}(T) & \mathrm{if \ unsaturated} \label{eq:dry_adiabat}\\ 
&= RT/L_\mathrm{i} & \mathrm{if \ saturated} \label{eq:moist_adiabat},
\end{eqnarray}\end{linenomath*}
with universal gas constant $R = 8.314$ J K$^{-1}$ mol$^{-1}$, ambient temperature $T$, and specific heat capacity $c_{p,i}$ and latent heat $L_{\mathrm{i}}$ per individual species $i$. We treat all gases as ideal to determine thermodynamic properties, such as their critical points and latent heats \citep{PierrehumbertBook2010}. Heat capacities are temperature-dependent:
\begin{linenomath*}\begin{eqnarray}
c_\mathrm{p,i}(\widehat{T}) = A_\mathrm{i} + B_\mathrm{i}\widehat{T} + C_\mathrm{i}\widehat{T}^2 +D_\mathrm{i}\widehat{T}^3 + E_\mathrm{i}/\widehat{T}^2,
\end{eqnarray}\end{linenomath*}
with $\widehat{T} = T(K) / 1000$, and the fit constants $A_\mathrm{i}$--$E_\mathrm{i}$ provided in the NIST Chemistry WebBook, SRD 69 \citep{cox1984codata,chase1998nist}.
\begin{figure}[hbt]
\centering
\includegraphics[width=0.49\textwidth]{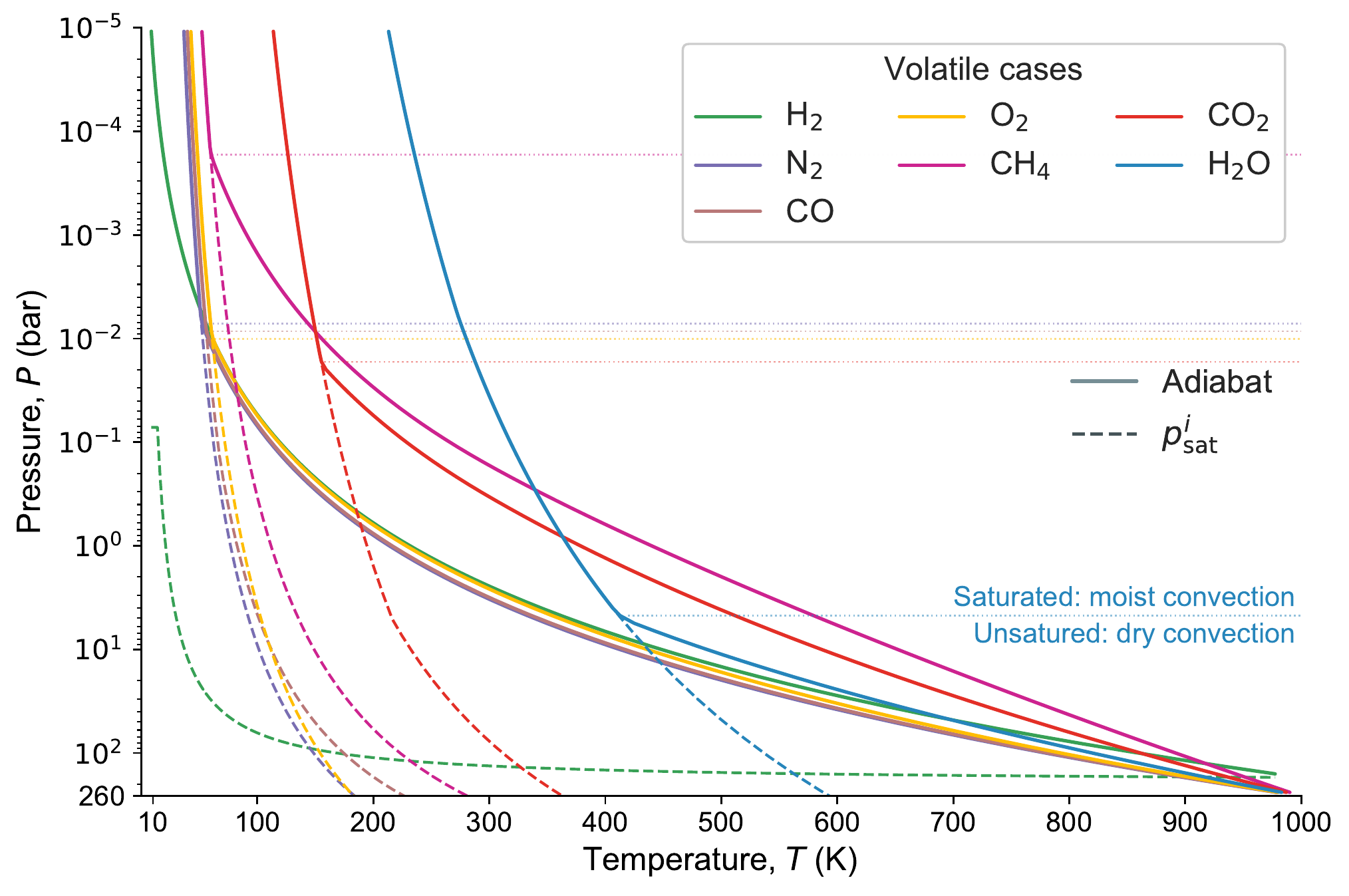}
\caption{\textsf{Thermal structure for single species atmospheres with fixed surface pressure equivalent $P_\mathrm{surf} = 260$ bar and surface temperature $T_\mathrm{surf} = 1000$ K. In the lower troposphere the atmospheric structure aligns with the dry adiabat until it intersects with the species-dependent saturation vapor curve, $p^i_\mathrm{sat}$, at which point condensates form and the lapse rate follows a moist convective profile (Eqs. \ref{eq:dry_adiabat} and \ref{eq:moist_adiabat}).}}
\label{fig:adiabat_example1}
\end{figure}
When condensation occurs during upward integration along the adiabat (Eq. \ref{eq:moist_adiabat}), a fraction of the gas phase is converted to a condensate phase according to the saturation vapor curve $p^i_\mathrm{sat}$ of the respective species and removed from the gas parcel \citep{PierrehumbertBook2010}. In the coupled framework, this may occur at ground level indicating formation of an ocean (or glacier) \citep{2017JGRE..122.1539M}. The formulation follows a dry adiabat in the lower troposphere, and a moist convective profile once the dry adiabat intersects the Clausius-Clapeyron slope where condensation occurs. The behavior of the adiabat structure is illustrated in Fig. \ref{fig:adiabat_example1} for H$_2$, H$_2$O, CO$_2$, CH$_4$, CO, O$_2$, and N$_2$, with surface pressure $P_\mathrm{surf} = 260$ bar, surface temperature $T_\mathrm{surf} = 1000$ K, and top-of-atmosphere (TOA) pressure of $10^{-5}$ bar = $1$ Pa.

To account for a potential stratosphere, we iterate between the adiabat adjustment and radiation atmosphere sub-modules by computing the atmospheric heating rate $\mathcal{H}$ (see below). We set the atmosphere to isothermal above the tropopause \citep{2016ApJ...829...63S} and recompute the radiation balance using this adjusted $T$--$P$ profile. This procedure yields an approximation to the tropopause location and stratospheric temperature structure and is computationally efficient compared to a time-stepped radiative equilibrium calculation. To compute the heat flux to space depending on composition and thermal profile of the atmosphere, we utilize the \textsc{socrates} radiative transfer suite \citep{edwards1996socrates,sun2011improving}. \textsc{socrates} is incorporated in the UK Met Office Unified Model \citep{2014AA...564A..59A,2016AA...595A..36A} and \textsc{rocke-3d} \citep{2017ApJS..231...12W}, and solves the plane-parallel, two-stream approximated radiative transfer equation, which we employ in the limit without scattering. The thermal flux is
\begin{linenomath*}\begin{eqnarray}
\pm \frac{1}{D} \frac{\mathrm{d} F^{\uparrow \downarrow}_{\mathrm{th}}}{\mathrm{d} \tau} = F^{\uparrow \downarrow}_{\mathrm{th}} - \pi B(T), \label{}
\end{eqnarray}\end{linenomath*}
with thermal upward and downward fluxes, $F^{\uparrow}_{\mathrm{th}}$ and $F^{\downarrow}_{\mathrm{th}}$ respectively, optical depth $\tau$, Planck intensity $B(T)$, temperature $T$, and diffusivity factor $D = 1.66$. Note that $F$, $\tau$, and $B$ are dependent on wavenumber $\tilde{\nu}$, which we omit here for clarity of notation. The incident stellar radiation is expressed as the direct, unscattered component,
\begin{linenomath*}\begin{eqnarray}
F_{\star}^{\downarrow} = F_{\star,\mathrm{TOA}} \cdot \mathrm{e}^{-\tau / \mu} / \mu,
\end{eqnarray}\end{linenomath*}
with $\mu = | \mathrm{cos} \, \theta |$ and mean stellar zenith angle $\theta = 54.55^{\circ}$, as a compromise between a daytime-averaged and insolation-weighted choice \citep{2014JAtS...71.2994C}.
The surface-averaged stellar flux at the top of the atmosphere (TOA) is calculated as
\begin{linenomath*}\begin{eqnarray}
F_{{\star},\mathrm{TOA}}(\tau_{\star}) = \frac{1-\alpha_\mathrm{p}}{4} \frac{\mathcal{L}_{\star}(\tau_{\star})}{4 \pi a_\mathrm{p}^2},
\end{eqnarray}\end{linenomath*}
with planetary albedo, bolometric luminosity of the star $\mathcal{L}_{\star}$, and mean orbital distance of the planet $a_\mathrm{p}$. $\tau_{\star}$ indicates the time since star formation. We assume a constant planetary albedo $\alpha_\mathrm{p} = 0.2$, which reasonably approximates the albedo over a wide temperature interval of steam atmospheres \citep{2015ApJ...806..216H}. Varying albedo due to scattering or cloud effects can influence the heat budget of the cooling planet \citep{2019Icar..317..583P}. For the time evolution of the stellar luminosity $\mathcal{L}_{\star}$ we employ a bilinar interpolation in the mass-time [$M_{\star}$, $\tau_{\star}$] plane of the stellar evolution models of \citet{1998AA...337..403B,2015AA...577A..42B} and \citet{2000AAS..141..371G}. Since we exclusively focus on Sun-like stars, we use the Solar spectrum from \citet{Kurucz1995}. The total upward and downward fluxes are the sum of the thermal and stellar components,
\begin{linenomath*}\begin{eqnarray}
F^{\uparrow \downarrow}_\mathrm{net}=F_{\mathrm{th}}^{\uparrow \downarrow}+F_{{\star}}^{\downarrow} \label{eq:thermal_stellar_flux}
\end{eqnarray}\end{linenomath*}
and thus the total atmospheric heat flux of the planet, which is required for the upper boundary condition of the interior, is
\begin{linenomath*}\begin{eqnarray}
F^{\uparrow}_\mathrm{MO} \equiv F^{\uparrow}_\mathrm{atm}=F^{\uparrow}_\mathrm{net}-F^{\downarrow}_\mathrm{net}. \label{eq:net_flux}
\end{eqnarray}\end{linenomath*}
The optical depth at a particular wavenumber in differential form is 
\begin{linenomath*}\begin{eqnarray}
\mathrm{d} \tau &=& -k(z) \mathrm{d} z=-k_{\rho}(z) \rho(z) \mathrm{d} z \nonumber \\
&=& -\sum \zeta_{i}(z) k_{\rho}^{i}(z) \rho(z) \mathrm{d} z,
\end{eqnarray}\end{linenomath*}
with mass mixing ratio $\zeta_{i}$ and mass absorption coefficient $k_{\rho}^{i}$ for volatile species $i$, height $z$, total mass absorption coefficient $k_{\rho}(z)$, and total absorption coefficient $k(z) = k_{\rho}(z)\rho(z)$. The heating rate is then
\begin{linenomath*}\begin{eqnarray}
\mathcal{H}=-\frac{\mathrm{d} F}{\mathrm{d} z}=\frac{g \overline{\mu}_\mathrm{atm} P}{R T} \frac{\mathrm{d} F_\mathrm{atm}}{\mathrm{d} P},
\end{eqnarray}\end{linenomath*}
with gravitational acceleration $g$ and mean molecular weight $\overline{\mu}_\mathrm{atm}$. \textsc{socrates} achieves computational efficiency by employing the correlated-$k$ method to solve the transmission function: absorption coefficients with similar values are grouped and summed over the atmospheric layers with different weights. $k$-coefficients are calculated using a combination of correlated-$k$ \citep{1989JQSRT..42..539G} and the exponential sum fitting of transmissions \citep{wiscombe1977exponential}. Line-by-line opacity data is banded into $n_{k}$ subintervals and an average absorption coefficient is calculated from the top of the atmosphere to an optical depth of one,
\begin{linenomath*}\begin{eqnarray}
k_{\rho, \mathrm{ avg }}^{i}(\tilde{\nu}) u_{\tau=1}^{i}&=&\int_{z_{z=1}}^{\infty} \mathrm{d} z^{\prime} \zeta_{i}\left(z^{\prime}\right) \rho\left(z^{\prime}\right) k_{\rho}^{i}\left(\tilde{\nu}, z^{\prime}\right) \nonumber \\
&=&\frac{1}{g} \int_{0}^{P_{\tau=1}} \mathrm{d} P^{\prime} \zeta_{i}\left(P^{\prime}\right) k_{\rho}^{i}\left(\nu, P^{\prime}\right) \equiv 1,
\end{eqnarray}\end{linenomath*}
with column density $u_{\tau=1}^{i}$ from the top of the atmosphere to $\tau = 1$ for species $i$. For each subinterval $l$ the weighted transmission from line-by-line coefficients is fitted to the transmission function, such that
\begin{linenomath*}\begin{eqnarray}
\int_{g_{l}}^{g_{l+1}} \mathrm{d} g \, w(g) \mathrm{e}^{k_{\rho}(g) u_{j}} \approx \mathrm{e}^{k_{\rho, \mathrm{opt}}^{l} u_{j}}
\end{eqnarray}\end{linenomath*}
with transmission $\mathrm{e}^{k_{\rho, \mathrm{opt}}^{l} u_{j}}$ for a set of $n_u$ column densities, $u_j$, weighting function $w$, and optimal $k$-coefficient $k_{\rho, \mathrm{opt}}^{l}$ in the given subinterval $l$. Here, we use differential Planckian weights $w_{l}$ at transmission temperature normalized per band,
\begin{linenomath*}\begin{eqnarray}
\int_{0}^{1} \mathrm{d} g \, w(g)=\sum_{l=1}^{n_{k}} w_{l}=1.
\end{eqnarray}\end{linenomath*}
The total flux per wavenumber band $b$ is then
\begin{linenomath*}\begin{eqnarray}
F_{b}=\sum_{l=1}^{n_{k}} w_{l}^{\tau=1} F_{l},
\end{eqnarray}\end{linenomath*}
with weights $w_{l}^{\tau=1}$, where the optical depth $\tau = 1$ \citep{2014AA...564A..59A}, and flux $F_{l}$ per subinterval. We use a band structure that covers the wavenumber range from 1--30000 cm$^{-1}$ and consists of 318 individual bands with a spacing of $\Delta \nu_{\mathrm{1-120}}$ = 25 cm$^{-1}$ from 1--3000 cm$^{-1}$, $\Delta \nu_{\mathrm{121-280}}$ = 50 cm$^{-1}$ from 3000--11000 cm$^{-1}$, and $\Delta \nu_{\mathrm{281-318}}$ = 500 cm$^{-1}$ from 11000--30000 cm$^{-1}$. We tabulate the $k$-coefficients on a $P$, $T$ grid of 21 pressures, $P \in$ [ $10^{-11}$, $10^{-8}$, $10^{-6}$, $10^{-5}$, $10^{-4}$, $5\times10^{-4}$, $10^{-3}$, $5\times10^{-3}$, $10^{-2}$, $5\times10^{-2}$, $10^{-1}$, $5\times10^{-1}$, $10^{0}$, $5\times10^{0}$, $10^{+1}$, $5\times10^{+1}$, $10^{+2}$, $5\times10^{+2}$, $10^{+3}$, $5\times10^{+3}$, $10^{+4}$ ] bar, and 15 temperatures, $T \in$ [ $100$, $250$, $400$, $600$, $800$, $1000$, $1250$, $1500$, $1750$, $2000$, $2333$, $2666$, $3000$, $3500$, $4000$ ] K. The opacity data that is used as input to generate the $k$ tables is referenced in Tab. \ref{tab:A1}.
\section{Results} \label{sec:results}
We explore the energetic feedback between a solidifying magma ocean and its outgassed atmosphere. By focusing on single species atmospheres consisting of either H$_2$, H$_2$O, CO$_2$, CH$_4$, CO, O$_2$, or N$_2$, we can isolate the influence of radiative properties and partitioning behavior on planetary cooling timescales. As a baseline for comparing the thermal chronologies of the coupled magma ocean--atmosphere systems, we choose the initial volatile abundance to produce an outgassed atmosphere of 260 bar after mantle solidification. We ignore trapped volatiles in the solidified rock since this abundance is negligible compared to volatile partitioning into the melt. For the case of H$_2$O, 260 bar roughly corresponds to a steam atmosphere equivalent of one Earth ocean. All models begin with a fully molten planetary mantle at 100 Myr after star formation.

\subsection{Steady-state atmospheric radiation balance}

\begin{figure*}[hbt]
\centering
\includegraphics[width=0.99\textwidth]{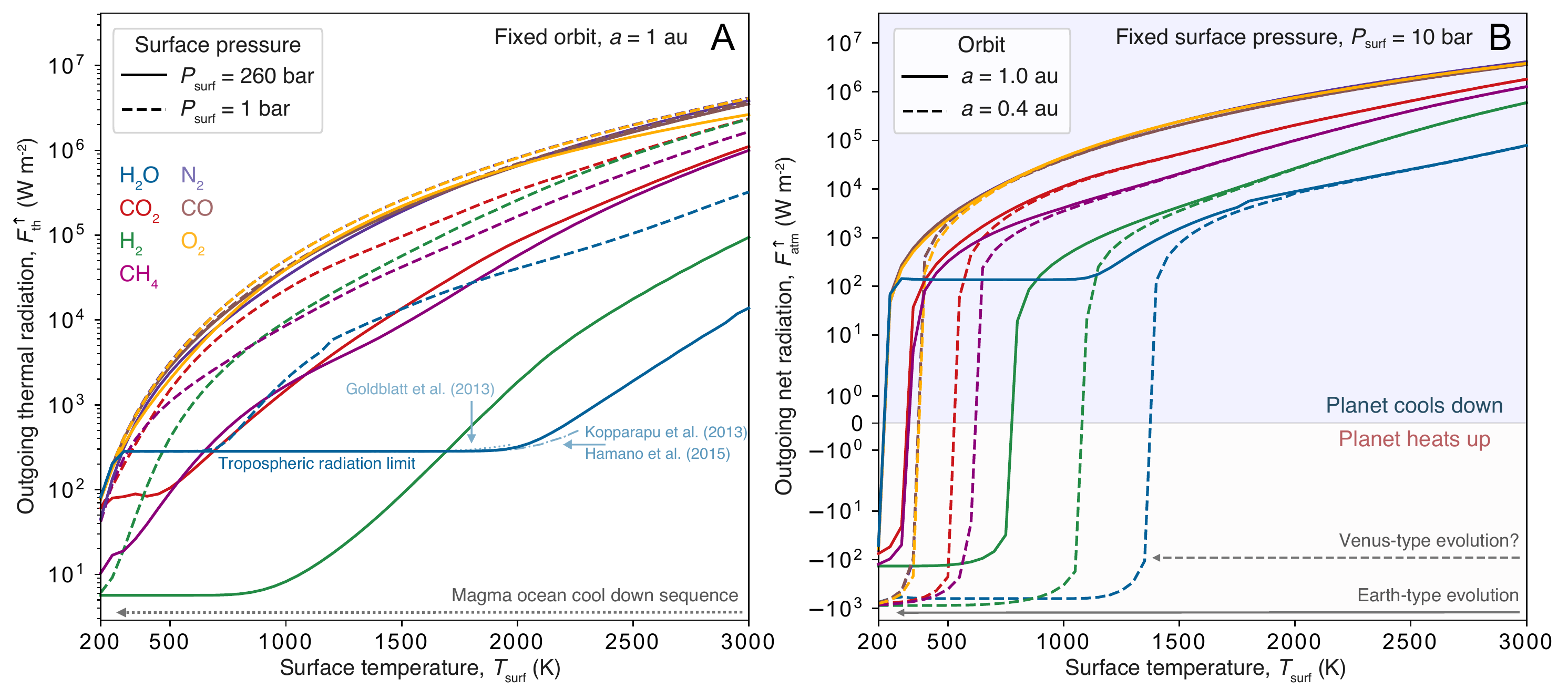}
\caption{\textsf{Outgoing radiation for single-species atmospheres and varying planetary settings. (\textbf{A}) Outgoing thermal radiation as a function of surface temperature for planetary orbital distance $a$ = 1 au, and surface pressure equivalent of $P_{\mathrm{surf}}$ = 260 bar or 1 bar. (\textbf{B}) Outgoing net radiation (thermal minus instellation, Eqs. \ref{eq:thermal_stellar_flux} and \ref{eq:net_flux}) as a function of surface temperature for $a$ = 1.0 au or $a$ = 0.4 au, with fixed surface pressure of $P_{\mathrm{surf}}$ = 10 bar and star age $\tau_{\star}$ = 100 Myr. For closer-in planets with fixed atmospheric content the incoming stellar radiation exceeds the heat flux to space and they enter the heating regime. The shown results are computed with a fixed planetary albedo of $\alpha_\mathrm{P} = 0.2$.}}
\label{fig:radiation_limits}
\end{figure*}

Before proceeding to evaluate the evolution of the coupled mantle-atmosphere system during magma ocean solidification, we illustrate the heat flux properties of the radiative-convective atmosphere model. We first discuss the steady-state behavior in order to validate our atmospheric model with previous approaches and illustrate key physical mechanisms that control the heat loss to space. In Fig. \ref{fig:radiation_limits}A we compare the outgoing thermal radiation ($F^{\uparrow}_\mathrm{th}$) for all volatiles for surface pressures of 260 bar and 1 bar, which capture the instantaneous state of the atmosphere near the end and start of outgassing, respectively. For the H$_2$O--260 bar case we recover the tropospheric radiation limit \citep{Simpson1929,Nakajima1992}, with a near-constant $F^{\uparrow}_\mathrm{th}$ of 282.5 W m$^{-2}$ in the window from $\approx$300--2000 K surface temperature. Our computed radiation limit closely aligns with recent estimates from \citet{2013NatGe...6..661G}, \citet{2013ApJ...765..131K}, and \citet{2015ApJ...806..216H} until about 2000 K, and further compares favorably to 280 W m$^{-2}$ in \citet{2017JGRE..122.1539M} and 282 W m$^{-2}$ in \citet{2019ApJ...875...31K}. For 260 bar, H$_2$ shows a strong depression in the outgoing radiation below the H$_2$O radiation limit for $T_{\mathrm{surf}} \lesssim 1700$ K which appears to be due to its strongly pressure-dependent opacity. However, validation of the opacity trend at the low temperature range near the surface where H$_2$ has high density requires a non-ideal H$_2$ equation of state. At the lower pressure of 1 bar the radiation limit for H$_2$O ranges from $T_{\mathrm{surf}} \approx$ 300--700 K and the outgoing radiation for H$_2$ is below H$_2$O for $T_{\mathrm{surf}} \lesssim 450$ K. CO$_2$ and CH$_4$ have similar $F^{\uparrow}_\mathrm{th}$ from $\approx$ 500 K and higher, which is intermediate compared to the lowest $F^{\uparrow}_\mathrm{th}$ (H$_2$) and the highest $F^{\uparrow}_\mathrm{th}$ (O$_2$). Relative to the other volatiles, N$_2$, CO, and O$_2$ do not significantly restrict the heat flux to space. None of the volatiles show strong depression in the outgoing radiation at 1 bar and high temperature.

Fig. \ref{fig:radiation_limits}B displays the net outgoing flux ($F^{\uparrow}_\mathrm{atm} \equiv F^{\uparrow}_\mathrm{MO}$, Eq. \ref{eq:net_flux}), which serves as the boundary condition at the interior--atmosphere interface when the surface is mostly molten. A significant feature of the 1.0 au cases is the extended radiation limits for H$_2$O between $T_{\mathrm{surf}} \approx$ 300--1000 K and for H$_2$ for $T_{\mathrm{surf}} \lesssim$ 750 K. Once planets are cool enough, their atmospheres become optically thick, and the outgoing radiation becomes smaller than the incoming stellar radiation. This transition occurs below 300 K except for H$_2$, so planets at 1.0 au with 10 bar H$_2$ atmospheres cannot cool down below $\approx$ 750 K in the absence of other processes, such as escape. For closer-in planets at 0.4 au, the transition points from the cooling to the heating regimes for the various volatiles are separated over a larger range of surface temperature than for planets at 1 au. CO$_2$ and CH$_4$ show net flux depressions at $T_{\mathrm{surf}} \lesssim$ 500 K, and CO$_2$ and CH$_4$ at $T_{\mathrm{surf}} \lesssim$ 300 K. The steady-state calculations reveal the influence of the planetary orbit and atmospheric composition on planetary heat loss, but they do not consider the time variations in both atmospheric volatile budget and thermal state due to progressive solidification of the mantle. The calculations suggest that all volatiles with significant flux depression (H$_2$O, H$_2$, CO$_2$, and CH$_4$) may experience a feedback between outgassing and solidification, resulting in marked differences in the cooling timescales. The timescales and atmospheric budgets, however, depend additionally on the solubility of the volatile in the melt. In the next section we thus consider the time-dependent feedbacks between the interior and atmosphere during the early evolution of planets starting from a fully-molten magma ocean.
\subsection{Time evolution of the coupled magma ocean--atmosphere}

\begin{figure*}[hbt]
\centering
\includegraphics[width=0.99\textwidth]{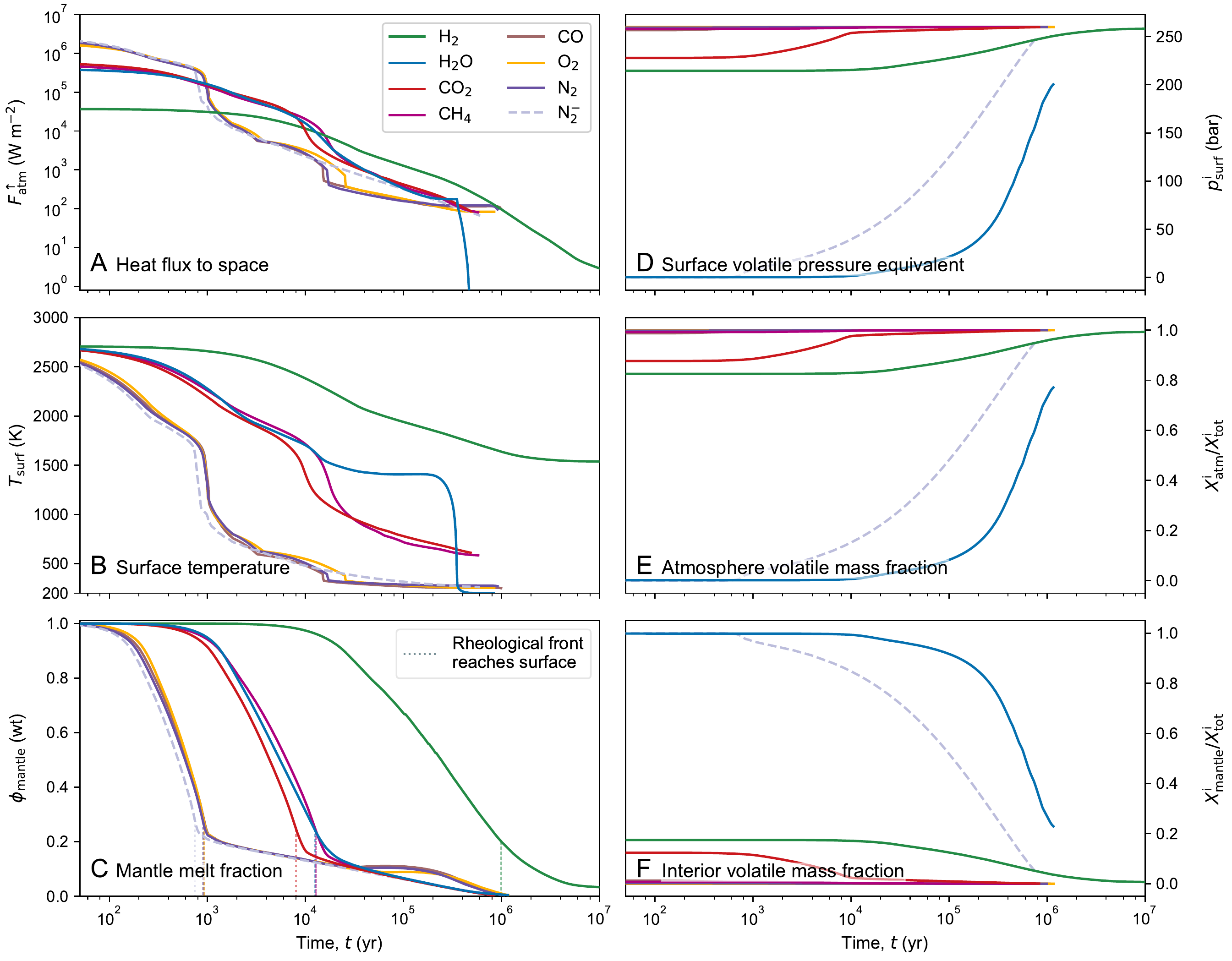}
\caption{\textsf{Coupled evolution of mantle and atmosphere. N$_2$ is considered for an oxidized (N$_2$) and reduced (N$_2^-$) case. Evolution can be classified into three groups of similar behavior: (i) H$_2$, with the most protracted solidification chronology, (ii) H$_2$O, CO$_2$, and CH$_4$ with intermediate cooling time, and (iii) CO, O$_2$, and N$_2$/N$_2^-$ with the shortest cooling time. Times shown are relative to model start, i.e., 100 Myr shifted to star formation. See main text for a more extensive description.}}
\label{fig:compare_global}
\end{figure*}

Fig. \ref{fig:compare_global} displays the evolution of a solidifying magma ocean for an atmosphere dominated by either H$_2$, H$_2$O, CO$_2$, CH$_4$, CO, O$_2$, or N$_2$~for an Earth-sized planet at 1.0 au from a Sun-like star after 100 Myr after star formation. Simulations terminate when the planet reaches a global melt fraction of $\phi_\mathrm{mantle} \leq 0.5$ wt.\%. The cooling sequence can be classified into three principal episodes, which vary in duration depending on the volatile species. The first phase is dominated by a fully molten magma ocean ($\phi_\mathrm{mantle} \approx$ 1) and varies from $\approx 10^2$ yr (CO, O$_2$, and N$_2$) to $\approx 10^3$ yr (H$_2$O, CO$_2$, CH$_4$), and $\approx 10^4$ yr (H$_2$) in duration. The interior temperatures are superliquidus and the interior convects vigorously which maintains a high surface temperature. During this phase, heat flux, surface temperature, and outgassing of volatiles to the atmosphere vary only minimally. Once the mantle starts to crystallize, the surface temperature decreases faster than during the fully molten phase because heat is not delivered as efficiently to the surface from the deep interior. This is because heat transport is diminished in mixed solid--liquid aggregates or fully solid rock relative to molten magma due to larger viscosity. The second phase occurs when the magma ocean is solidifying from the core-mantle boundary to the surface. This phase lasts until $\approx 10^3$ yr for CO, O$_2$, N$_2$, and $\approx 10^4$ yr for H$_2$O, CO$_2$, CH$_4$, and $\approx 10^6$ yr for H$_2$. Subsequently the rheological front, which is defined as the (quite abrupt) increase in viscosity from a melt- to solid-dominated aggregate ($\phi \approx 0.4$), reaches the surface. This signals a transition to sluggish convection in the interior and hence a protracted cooling phase (phase 3). Solidification proceeds until the mantle reaches near-zero melt fraction, which takes $\approx$ 1 Myr for all volatiles but H$_2$. The H$_2$ planet case does not solidify within 100 Myr simulation runtime.

The heat flux, surface temperature, and mantle fraction (Fig. \ref{fig:compare_global}A, B, and C) result from the feedback between mantle solidification and thermal blanketing by the atmosphere, which limits the heat flux to space depending on the atmospheric conditions. Fig. \ref{fig:compare_global}D, E, F show the outgassing sequence that defines the surface pressure, and total atmospheric and interior volatile mass fraction. The outgassing sequences differ substantially between cases due to the different solubilities of the volatiles in melt (Fig. \ref{fig:solubilities}). Generally, H$_2$O and N$_2^-$ are the most soluble in magma and hence they outgas late relative to other cases. H$_2$O does not outgas completely to establish a 260 bar atmosphere because its solubility is high enough to trap a fraction of H$_2$O in the remaining $\approx$ 0.5 wt.\% of melt. O$_2$ is treated as insoluble, and hence its atmospheric mass and surface pressure is constant. For all other cases, outgassing is effectively driven by the reduction in mantle melt fraction, hence outgassing mostly occurs during phase 2 when solidification begins until the rheological front reaches to the surface (at which time $\approx$70--80 wt.\% of the planetary mantle is solidified). Outgassing then proceeds until all the volatile mass is transferred to the atmosphere from the interior. Even though volatiles such as H$_2$O can be stored in nominally anhydrous minerals, their abundance in the deep mantle is typically around $\sim 0.1$ wt.\% \citep{PeslierISSI2018} which is negligible in comparison to the atmospheric reservoir.

\begin{figure}[h!]
\centering
\includegraphics[width=0.49\textwidth]{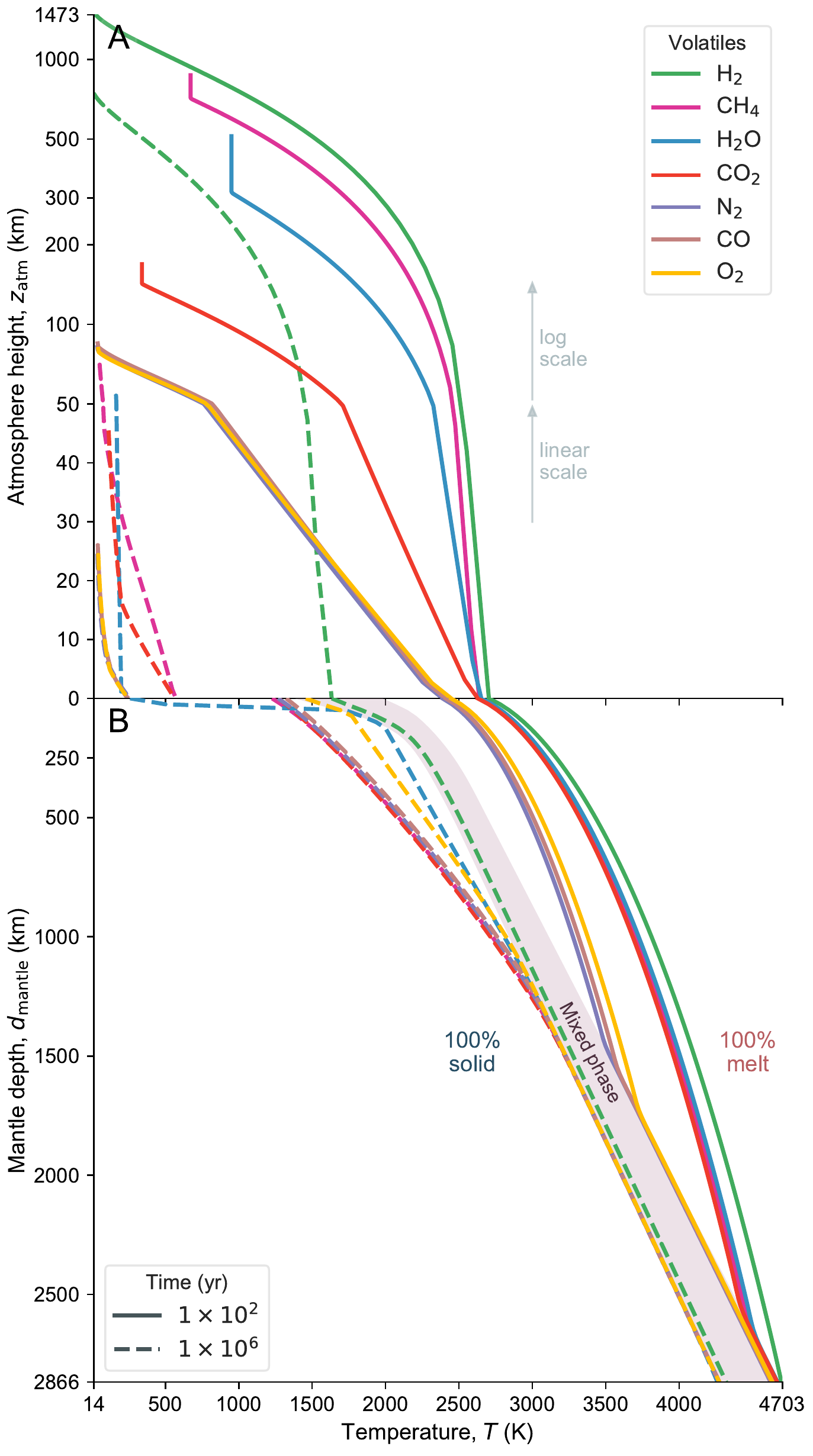}
\caption{\textsf{Thermal and vertical structure of the atmosphere and interior at two snapshots during evolution. (\textbf{A}) Atmospheric height stratification. Early atmospheres are thermally expanded and lower molecular weight atmospheres extend to great height. Note the transition in the height axis between linear and log scale at 50 km. (\textbf{B}) Interior evolution. The transition region from fully-molten to fully-solidified rock is illustrated with a gray background ('Mixed phase'). Right of this area heat transfer is very rapid and dominated by vigorous convection in the liquid magma, left of it heat transfer is much slower and dominated by solid-state creep processes.}}
\label{fig:compare_stacked}
\end{figure}

To illustrate the combined evolution of the planet's interior and atmosphere we show each case at two distinct times in Fig. \ref{fig:compare_stacked}. This plot shows the pressure--temperature structure of the atmosphere due to both its composition and radiative-convective structure. H$_2$ displays the most pronounced vertical stratification because of its low molecular weight and slow cooling, followed by CH$_4$, H$_2$O, CO$_2$, then CO, N$_2$, and CO. However, at 1 Myr the H$_2$O case shows departure of its vertical structure relative to the earlier time and the CH$_4$ and CO$_2$ cases, and its surface temperature is closer to the ones of CO, O$_2$, and N$_2$. This emphasizes the influence of the coupled thermal evolution of the atmosphere on the overall planetary structure. Fig. \ref{fig:compare_global}B reveals the distinct evolution when comparing H$_2$O with the CH$_4$ and CO$_2$ cases. The surface temperature for the H$_2$O case is near-constant between $10^4$--$10^6$ yr. However, once the heat flux from the interior drops significantly the surface temperature falls rapidly (Fig. \ref{fig:compare_global}A) and the atmosphere collapses. This is because for intermediate surface temperatures the atmospheric heat flux remains constant since the upper atmosphere aligns with the H$_2$O moist adiabat, i.e., the heat flux is regulated by the radiation limit (Figs. \ref{fig:adiabat_example1} and \ref{fig:radiation_limits}). Hence, when the internal heat flux cannot sustain heat loss above the radiation limit anymore, the lithosphere cools rapidly until equilibrium between the internal and atmospheric heat fluxes is reached. Because of the high solubility of H$_2$O, most H$_2$O outgasses after $\gtrsim$90 wt.\% of the mantle is already solidified and the surface temperature favors condensation. Hence outgassed water immediately forms a condensed surface water reservoir rather than contributing to the atmospheric greenhouse effect. CO$_2$ and CH$_4$ on the other hand are less effective as absorbers and the planet can cool faster until $\approx$1 Myr after model start. These volatiles, however, are significantly less soluble in magma compared to H$_2$O , and do not facilitate a condensation-induced radiation limit for temperatures relevant to magma ocean cooling (Fig. \ref{fig:radiation_limits}). Therefore, the corresponding planets steadily cool down and their atmospheres grow to their full outgassed 260 bar surface pressure early during magma ocean evolution.

\subsection{Observational links between atmospheric and interior state}

\begin{figure*}[htb]
\centering
\includegraphics[width=0.99\textwidth]{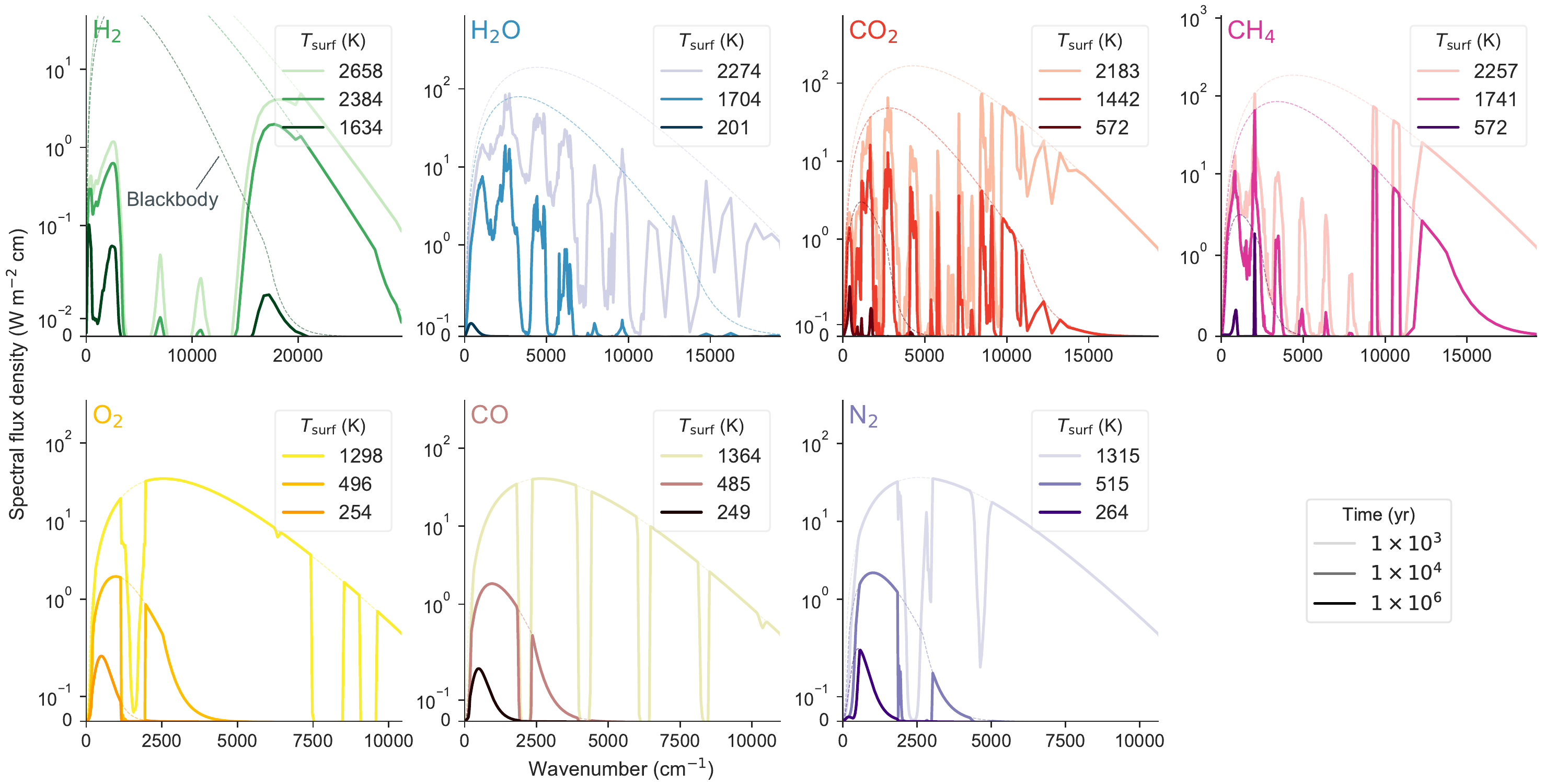}
\caption{\textsf{Top-of-atmosphere net spectral flux relative to their blackbody spectra for $t=$ $10^3$, $10^4$, and $10^6$ yr during magma ocean evolution. These times correspond to when the rheological transition reaches the surface in the different volatile classes (Fig. \ref{fig:compare_global}).}}
\label{fig:compare_atm}
\end{figure*}

The sequence of planet solidification for the different volatiles produces time-evolving atmospheres with different thermal states and spatial extents, which furthermore can provide predictions on the observable spectral signature of the atmosphere. For single-species atmospheres, the top-of-atmosphere emission spectra show distinct features during their evolution and are distinguishable from one another. Fig. \ref{fig:compare_atm} displays the planetary spectra at three distinct times during evolution when the rheological transition reaches the mantle--atmosphere interface for the different volatile groups at $t=$ $10^3$, $10^4$, and $10^6$ yr (Fig. \ref{fig:compare_global}). The rheological transition denotes when the mode of heat transfer in the mantle transitions from vigorous convection and rapid cooling to more protracted cooling that is dominated by solid-state convection. In general, higher wavenumber features are muted as the planet solidifies and the spectra across the different absorbers reveal distinct features from each another. H$_2$ during early times is dominated by the wavenumber regions around $\nu \approx$ 20000 cm$^{-1}$ but at 1 Myr these regions do not contribute significantly to the heat flux anymore and the planet is dominantly cooling through the regions around $\nu \lesssim$ 3000 cm$^{-1}$. Similar trends occur for H$_2$O, CO$_2$, and CH$_4$. H$_2$O evolves in regions of 4000--6000 cm$^{-1}$, CO$_2$ in several distinct regions from $\approx$4000--15000 cm$^{-1}$, and CH$_4$ at $\approx$3000 cm$^{-1}$ and between $\approx$9000--11000 cm$^{-1}$. CO, N$_2$, and O$_2$ cool very rapidly and their absorbing regions are confined to ranges around $\approx$1500 cm$^{-1}$ for O$_2$, $\approx$2000 cm$^{-1}$ and $\approx$4100 cm$^{-1}$ for CO, and $\approx$2500 cm$^{-1}$ and $\approx$4500 cm$^{-1}$ for N$_2$. In summary, H$_2$, H$_2$O, CO$_2$, and CH$_4$ display the strongest departures from the blackbody curve, which means they absorb more longwave radiation that is emitted from the planetary surface. O$_2$, CO, and N$_2$ bear significant resemblance to the blackbody curve, so alter the radiation properties of the planet to a lesser degree. Nevertheless, N$_2$ displays absorption features around $\approx$2500 and $\approx$4700 cm$^{-1}$ during the fully-molten phase.

\begin{figure*}[htb]
\centering
\includegraphics[width=0.99\textwidth]{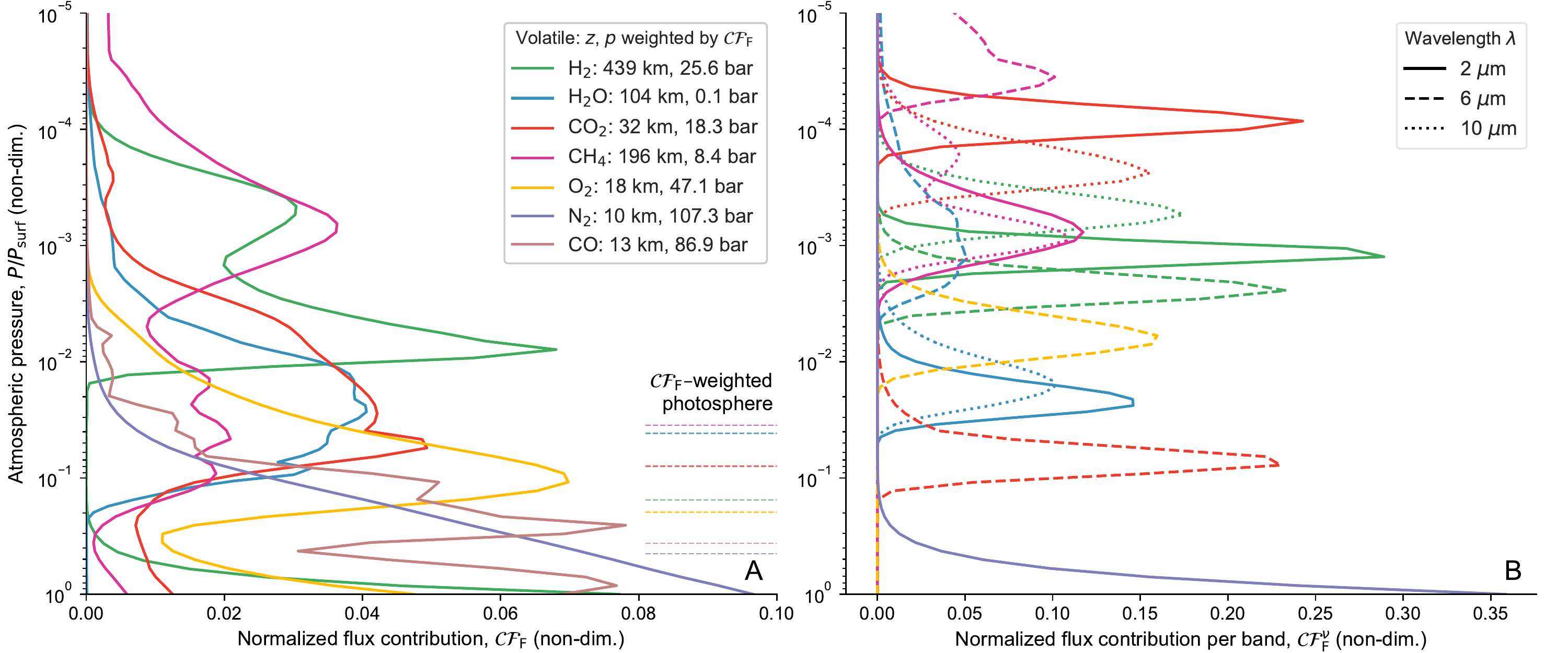}
\caption{\textsf{Total and band flux contribution from different atmospheric levels when the rheological front reaches the surface for each simulation (cf. Fig. \ref{fig:compare_global}). (\textbf{A}) Normalized total flux contribution. Additionally indicated are the $\mathcal{CF}_\mathrm{F}$--weighted pressure levels of the planets (\emph{photosphere}). (\textbf{B}) Flux contribution for wavelength bands at $\lambda$ = 2, 6, and 10 $\mu\mathrm{m}$.}}
\label{fig:compare_cff}
\end{figure*}

Astronomical observations are typically only possible at confined wavelength regions and thus it is necessary to connect magma oceans with outgassed atmospheres of different compositions to specific features at certain wavelengths. Therefore, Fig. \ref{fig:compare_cff} compares the contribution of different atmospheric layers to the outgoing flux for the total and wavelength-specific flux. We quantify the normalized fraction of outgoing net flux that escapes to space from a particular layer $j$ in the atmosphere \citep{2018ApJ...869...28D}: 
\begin{linenomath*}\begin{eqnarray}
\mathcal{CF}_\mathrm{F} = \frac{2 \pi B (\tau_{j,j+1})}{D} \left( e^{-D\tau_j} - e^{-D\tau_{j+1}} \right).
\end{eqnarray}\end{linenomath*}
Fig. \ref{fig:compare_cff}A illustrates the contribution of the top-of-atmosphere net outgoing flux from different atmospheric layers at different times, and visualizes the dominant photospheres of the magma ocean planets at the transition from melt- to solid-dominated. The flux for the CO, N$_2$, and O$_2$ cases is dominated by near-surface layers since these volatiles are least effective as absorbers. In particular for the N$_2$ case, the magma ocean surface itself is the dominant emitter, suggesting that planets with N$_2$ atmospheres may enable their magma ocean surface to be probed directly. Depending on atmospheric conditions (pressure--temperature), the photosphere shifts to different levels and can sometimes be vertically split, i.e., emission can be dominated by two or multiple distinct regions. For instance, the H$_2$ case displays two local maxima of the contribution function at $\approx 8 \times 10^{-3}$ and $\approx 1$ normalized pressure. The $\mathcal{CF}_\mathrm{F}$--weighted pressure level (the \emph{photosphere} of the planet) lies approximately between these two maxima. The large scale height in Fig. \ref{fig:compare_stacked} suggests that \textsc{socrates} overestimates the opacity of the upper of the two maxima since the height variation of gravity $g(z)$ is not considered. This could affect the cooling rate by underestimating the mass path, $dm \sim dp/g$. For a related discussion for water worlds see \citet{2019ApJ...881...60A}. Therefore, further work is required to estimate the cooling behavior of extended H$_2$ atmospheres. The H$_2$O and CO$_2$ cases mostly emit from intermediate pressure levels between $\approx 1 \times 10^{-1}$ and $\approx 5 \times 10^{-3}$ normalized pressure. This analysis demonstrates that we may be able to probe different layers in an atmosphere above a magma ocean (even down to the magma ocean surface) by observing planets with protoatmospheres composed of different dominant volatiles. Wavelengths exhibit different sensitivities with respect to their emitting photospheres. Fig. \ref{fig:compare_cff}B illustrates how the primary emitting surface shifts for 2, 6, and 10 $\mu\mathrm{m}$. For CO$_2$, 2 and 10 $\mu\mathrm{m}$ dominantly probe the upper atmospheric layers, but at 6 $\mu\mathrm{m}$ deep atmospheric layers contribute most significantly to the top-of-atmosphere flux. The photosphere depth is not a linear function of wavelength but intimately tied to the opacity of the volatile and the state of the magma ocean and atmosphere. Thus, observations that exploit multiple wavelengths can uncover the thermal stratification with atmospheric depths and reveal deeper or shallower parts of the temperature profile.
\begin{figure*}[htb]
\centering
\includegraphics[width=0.99\textwidth]{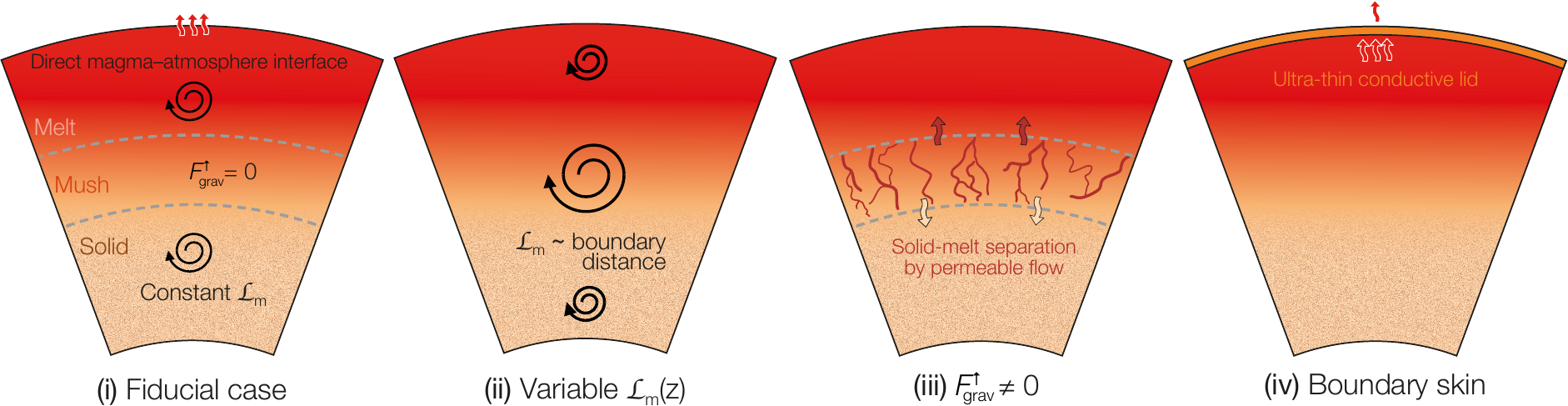}
\caption{\textsf{Illustration of interior model parameterizations and influence on mantle solidification. (\textbf{i}) The fiducial case with constant mixing length approximates 0-D approaches based on boundary layer theory and no separation of melt and solid occurs. Other cases relax an assumption: (\textbf{ii}) Mixing length $\mathcal{L}_{\mathrm{m}}$ scales with distance to the nearest domain boundary (magma ocean--atmosphere interface or core-mantle boundary), meaning that convection is most vigorous in the mid-mantle and decays towards the boundaries. (\textbf{iii}) Solid and melt can separate via permeable flow. (\textbf{iv}) The magma ocean and atmosphere are separated by an ultra-thin ($\sim$cm scale) conductive boundary skin, which reduces the surface temperature relative to the interior adiabat thereby increasing the cooling time.}}
\label{fig:interior_params}
\end{figure*}
\begin{figure*}[bt]
\centering
\includegraphics[width=0.99\textwidth]{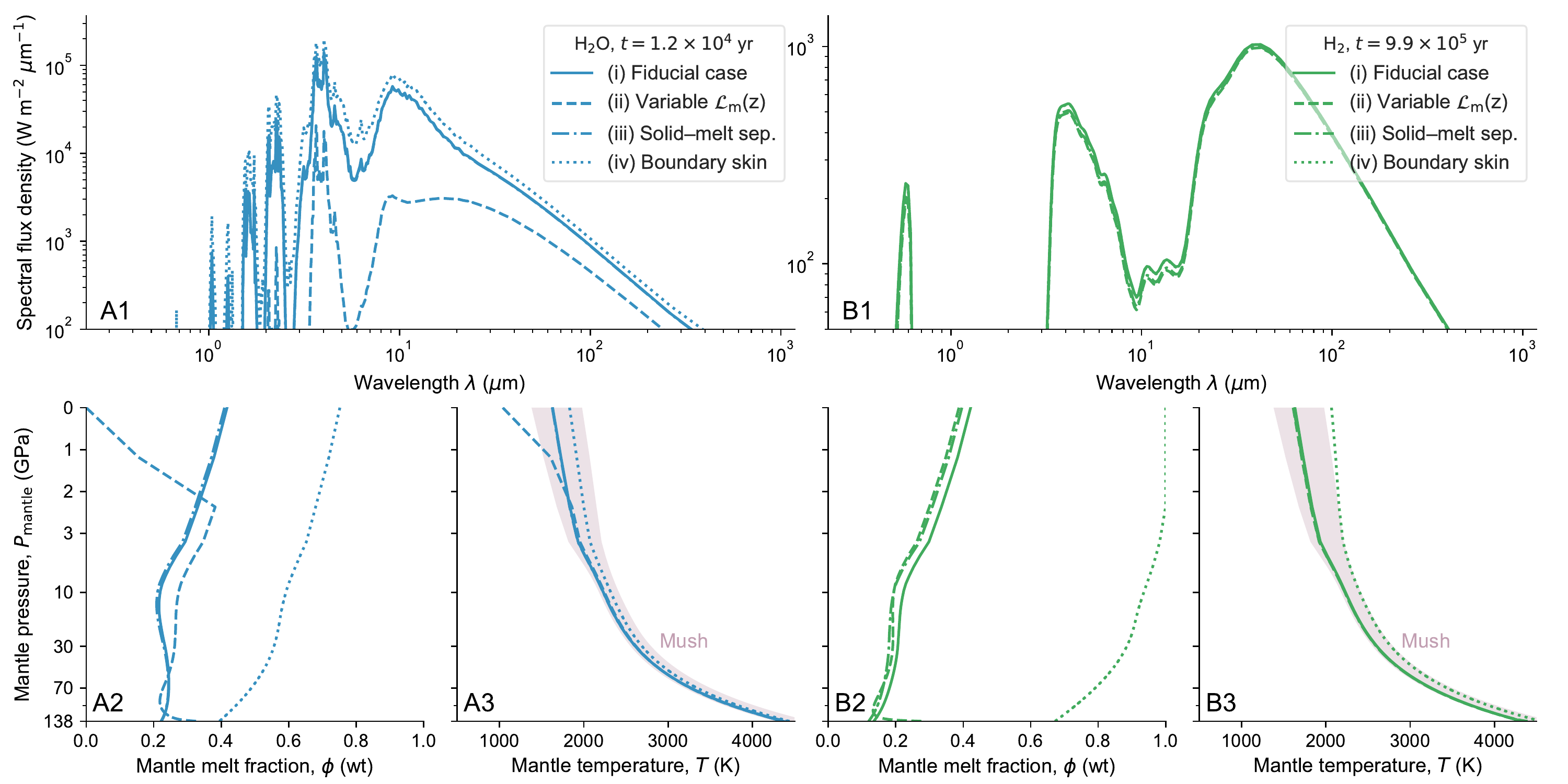}
\caption{\textsf{The influence of interior processes on spectral flux density (compare with Fig. \ref{fig:interior_params}), mantle melt fraction, and mantle temperature for the H$_2$O  and H$_2$ cases when the rheological transition reaches the surface in the fiducial case (Fig. \ref{fig:compare_global}). (\textbf{A}) H$_2$O case at $t \approx 10^4$ yr, and (\textbf{B}) H$_2$ case at $t \approx 10^6$ yr. In A3 and B3 the gray area denotes the mixed region of melt and solid. Note the non-linear y-axis scale in plots A2, A3 and B2, B3. The boundary skin (iv) delays cooling and modulates the brightness temperature of a largely molten planet relative to the fiducial case.}}
\label{fig:compare_int_setting}
\end{figure*}

Figures \ref{fig:interior_params} and \ref{fig:compare_int_setting} explore the sensitivity of the atmospheric state to processes that dictate the evolution of the planetary interior for the H$_2$O and H$_2$ cases. Fig. \ref{fig:interior_params} shows the parameter variations we consider and illustrates how these impact evolution of the interior (see Sect. \ref{sec:disc_interior} for discussion). Figure~\ref{fig:compare_int_setting} shows how the atmospheric state is perturbed based on processes operating in the interior. In general, spectral flux density variations, relative to the fiducial case, are more apparent for the H$_2$O case than the H$_2$ case. Varying the mixing length $\mathcal{L}_\mathrm{m}$ affects the vigor of convection in the mantle and hence leads to the strongest deviations. Solid--melt separation displays no considerable effect and inclusion of a boundary skin at the surface shows constant but small deviations by delaying cooling. An obvious feature for the H$_2$O case is the reduced melt fraction at the near-surface between $\approx$0--2.5 GPa for variable $\mathcal{L}_\mathrm{m}$ (ii). This melt depletion of the near-surface comes with decreased surface temperature and thus muting of spectral features, most notably between $\approx$5--20 $\mu m$ and up to two orders of magnitude at $\approx$6 $\mu m$. At a given time, the boundary skin (iv) leads to strong enhancement of mantle melt fraction relative to the fiducial case by delaying cooling. Thus, the spectrum for this case is shifted relative to the fiducial case at a roughly similar value over all wavelengths. For the H$_2$ case, the difference in the spectrum is limited to the variable mixing length. Even though the total mantle melt fraction for the boundary skin case deviates even stronger from the fiducial case (relative to the H$_2$O deviation), there is no observable difference in the atmospheric spectrum. This is due to the strong influence of the H$_2$ atmosphere, which dominates the cooling rate of the planet (Fig. \ref{fig:compare_global}). The most notable difference between the H$_2$O and H$_2$ cases is the lower mantle melt fraction for the variable mixing length parameterization, which is depleted in melt in the H$_2$ case. In addition, the H$_2$ case with the boundary skin remains above the solidus temperature, such that the upper mantle can sustain a surface magma ocean for an extended duration relative to the fiducial case and the H$_2$O case. In short, processes operating in the interior can affect the spectral properties of the overlying atmosphere by up to two orders of magnitude. These are reflected by changes of the spectrum shape, features at specific wavelengths, and whole-spectrum displacements according to Wien's law.

\section{Discussion} 
\label{sec:discussion}
Developing the theory and models to connect astrophysical observations to the atmospheric and interior state of terrestrial planets is crucial for elevating our understanding of rocky worlds within and beyond the Solar System. In this work we established a novel modeling framework to investigate the thermal and compositional evolution of young, rocky planets to further remove Solar System bias from considerations of early planetary evolution. In conjunction with astronomical reconnaissance of extrasolar planetary systems, this will allow to untangle the dominant physical and chemical mechanisms shaping the evolution of rocky planets from their formation to their long-term evolution. To this end, we focused on the radiative effects of varying dominant absorbers during magma ocean solidification.

\subsection{Atmosphere formation \& speciation}
\label{sec:disc_atm}
The bulk composition and eventual climate of evolved planets is a result of the physical and chemical mechanisms that operate during accretion and magma ocean evolution. We presented a modeling framework that vertically resolves the thermal structure of the planet from the core-mantle boundary at the base of the mantle to the top of the atmosphere. This will allow us in future to investigate vertical variations in mantle and atmospheric chemistry. We focused on the radiative effects of single-species atmospheres and simulated outgassing and progressing solidification, thereby establishing our coupled modeling approach as a novel tool to explore the early evolution of rocky planets. Our results highlight that various classes of cooling behavior can be expected for different volatiles. However, they do not elucidate the behavior of more complex speciation that may result from changes in the fractionation pattern during volatile delivery. From that perspective, the outgassing behavior of the two contrasting N$_2$ cases is particularly relevant, as it highlights the difference between reduced and oxidized planetary interiors. This emphasizes that redox evolution of the planetary mantle and atmospheric speciation are tightly interconnected, as suggested by laboratory experiments \citep{2020NatCo..11.2007D,2020GeCoA.280..281G}. Considering more realistic outgassing speciation as a result of the rock composition \citep{2017ApJ...843..120S,2018RvMG...84..393S,2020AA...636A..71H} is thus necessary. In addition, we here ignore the potential effects of dynamic trapping of volatiles due to inefficient melt drainage out of the freezing front during solidification \citep{2017GGG....18.3078H}. By altering the outgassing pressure, enhanced volatile trapping may influence the radiating level of the planet, but is mostly confined to the early stages of rapid freezing.

The strong effects of H$_2$ on the thermal sequence and crystallization are striking, even with the aforementioned caveat that part of the upper atmosphere opacity may be overestimated in our radiative transfer calculations. To compare the different absorbers using a similar standard, we considered a thick primordial atmosphere that strongly inhibits cooling. We do not consider loss by atmospheric escape \citep{2018AARv..26....2L,2019AREPS..47...67O} and thus cannot constrain the competition between outgassing and escape. However, except for the H$_2$ case the short magma ocean cooling times do not allow significant time for atmospheric loss. For long-lived magma oceans, atmospheric escape could play a crucial role during the magma ocean stage of planetary evolution. Recent studies exploring the super-Earth to mini-Neptune transition argue for a substantial influence of H$_2$ on the mass--radius relation due to its non-ideal partitioning into the planetary interior \citep{2018ApJ...854...21C,2019ApJ...887L..33K}. Whether or not Earth-sized planets in temperate orbits can acquire substantial H$_2$ atmospheres remains debated, but this could alter the planetary radius if the interior remains molten due to the difference in density between molten and solid rock \citep{2019AA...631A.103B}. Probing this transition with upcoming space missions will inform whether young rocky planets experience long-lived phases of melt--solid--atmosphere interaction as we predict for the H$_2$ case. This will influence whether the magma ocean stage can completely overprint prior compositional variation inherited during accretion \citep{2019NatAs...3..307L}.

\subsection{Geophysical history of rocky exoplanets}
\label{sec:disc_hitory}

Upcoming space missions and potentially ground-based surveys with infrared capabilities may probe the Earth-sized planetary regime on temperate orbits and will become primary tools to explore the composition and evolution of magma oceans \citep{2014ApJ...784...27L,2015ApJ...806..216H,2019AA...621A.125B}. We illustrate the intimate connection between interior state and atmospheric spectra, and reveal how probing different planets of different compositions can yield insights into interior processes. Planets that are dominated by atmospheric species such as N$_2$ or CO, or completely stripped of any atmosphere, may allow the direct observation of the planetary surface. Alternatively, the dependence of the planetary photosphere on composition and wavelength demonstrates that multi-wavelength observations may also provide opportunity to probe the temperature structure and atmospheric composition. Observations of magma oceans at different ages can reveal distinct evolutionary stages. Observations of planets around young stars can probe the fully-molten stage whereas observations of older planets reveals the sluggish phase of partial solidification; together these establish a chronology of planetary evolution. The short duration of the Earth-like cooling regime may limit the number of observational targets, depending on the age of the host star and volatile budget and composition \citep{2019AA...621A.125B}. However, planets inside the runaway greenhouse limit could offer plenty of targets with long-lived magma oceans and a wide variety of volatile inventories.

Probing the early stages of planetary evolution is relevant for the history of reduced and oxidized atmospheres. Photolysis and subsequent hydrogen escape could be a main mechanism for water depletion and oxygen build-up in rocky planetary atmospheres \citep{2015AsBio..15..119L,2016ApJ...829...63S}, and is a popular explanation to interpret exoplanetary observations in the context of planetary evolution and atmospheric composition \citep{2019Natur.573...87K}. However, our results illustrate that different atmospheric absorbers can produce a variety of thermal and compositional states that evolve as a planet cools, hence setting the stage for prebiotic chemical environments on young planets \citep{2012Icar..219..267W,2020SciA....6.3419S}. A natural next step is to incorporate and compare the influence of additional chemical and physical processes, such as atmospheric escape \citep{2018Icar..307..327O,2018AJ....155..195W,2020arXiv200711006R}, chemistry \citep{2012ApJ...761..166H,Pearce2020}, and condensates \citep{2017JGRE..122.1539M,2019Icar..317..583P}.

\subsection{Interior state \& tectonics}
\label{sec:disc_interior}

Generally speaking, we find that for all but H$_2$, magma ocean cooling in the Earth-type regime undergoes similar evolutionary stages as the canonical steam atmosphere: initial rapid cooling during the fully molten phase followed by a phase of more protracted cooling once the mantle reaches the rheological transition. The H$_2$ case, however, shows a distinct evolution with comparably protracted cooling and smoothing of the three distinct stages of solidification. In general, the different volatile cases differ by orders of magnitude with respect to the timescale of mantle crystallization. The solidification time has consequences for the compositional stratification of the post-magma ocean mantle mineralogy and style of solid-state convection \citep{elkins2008linked,2015JGRB..120.6085B,solomatov2015treatise}. Fast cooling, as observed for planets with O$_2$, CO, and N$_2$ atmospheres, promotes batch crystallization of planetary mantles which decreases the tendency for compositional stratification and large-scale overturn. In contrast, more protracted cooling, which is favored for H$_2$, H$_2$O, CO$_2$, and CH$_4$ cases, promotes fractional crystallization of cumulates and thus whole mantle overturn and compositional stratification.

Furthermore, the solidification time influences the onset of solid-state convection \citep{2017GGG....18.2785B,2017JGRE..122..577M} and the intrusion efficiency of upwelling magma columns, thereby regulating the post-magma ocean cooling regime of the planet \citep{2018NatGe..11..322L}. Tidal effects due to rapid rotation following giant impacts, however, can alter the energetic balance and magmatic evolution of planetary bodies \citep{2015EPSL.427...74Z,2020EPSL.53015885L}. From a long-term geodynamical perspective, mineralogical stratification and potential mantle overturn may result in different styles of mantle convection, separating whole-mantle convection from double-layered convection regimes with consequences for long-term outgassing behavior \citep{2018RSPTA.37680109S,2020arXiv200709021S}. In addition, we directly link the observed spectra of magma ocean planets to interior state by exploring the influence of interior processes on observables. For example, variations in the vigor of magma ocean convection are parameterized by the mixing length parameter, which is analogous to the eddy diffusivity $k_{zz}$ in atmospheric sciences. Such modeling choices cannot be constrained by observations in the Solar System since magma oceans are no longer present, and laboratory experiments are inherently limited in parameter space. Therefore, astronomical reconnaissance of magma ocean planets may help to further constrain the relative importance of energy transfer processes operating in planetary interiors and inform appropriate parameterizations.

\section{Summary \& conclusions} \label{sec:conclusions}

We investigated the evolution of the interior and atmosphere of rocky planets during magma ocean solidification with a focus on the radiative effects of protoatmospheres of different composition. The newly established framework of a radiative-convective atmospheric model coupled to a vertically resolved energy balance model of the planetary mantle allows us to constrain the thermo-physical evolution of magma ocean planets from the core-mantle boundary to the top of the atmosphere. We used the radiative-convective atmosphere model to demonstrate the volatile-specific planetary cooling rate as a function of surface temperature for fixed atmospheric pressures. We then used the coupled framework to model the time-dependence of magma ocean solidification for rocky planets dominated by atmospheres comprised of H$_2$, H$_2$O, CO$_2$, CH$_4$, CO, O$_2$, or N$_2$, finding strong variation in their thermal evolution and vertical stratification. We quantified the evolution of their atmospheric spectra and the heat flux contribution from different atmospheric levels during the evolution. This allows inferences of the interior mantle state from atmospheric signals extracted from astronomical observations. Finally, we demonstrated that the relative efficiency of energy transport by different interior processes may have a measurable effect on the associated protoatmospheres. Our coupled magma ocean--atmosphere framework thus expands the solution space in the spatial and compositional domain and makes it possible to investigate planet formation and evolution pathways that differ from the terrestrial planets in the Solar System.

Investigating magma ocean evolution with the coupled model, we find:

\begin{itemize}
  \item The thermal sequence of magma ocean solidification is influenced by the radiatively dominant volatile to first order; cooling can be protracted by several orders of magnitude for the same orbit and planet configuration. Ordered by increasing timescale of solidification, the volatiles are grouped into three classes:
  \begin{itemize}
    \item O$_2$, N$_2$, and CO are least effective as atmospheric absorbers and hence their magma oceans cool the fastest. Because of their low solubilities in silicate melts, their atmospheres form quickly to their full extent and remain at near-constant surface pressure due to their low condensation temperatures.
    \item H$_2$O, CO$_2$, and CH$_4$ establish an intermediate class with more protracted magma ocean cooling. Even though the tropospheric radiation limit for the H$_2$O case is most pronounced, CO$_2$ and CH$_4$ show similar solidification timescales. The reason for this is a combination of their different solubilities in magma, atmospheric extinction efficiencies, and condensation temperatures.
    \item H$_2$ shows the strongest effect on extending the duration of the magma ocean phase. Even though the evolution still charts the main phases of magma ocean cooling from fully molten to partially solidified to fully solidified, the transitions between these phases are smoothed and delayed relative to the other volatiles. Solidification with an H$_2$ atmosphere is not completed within 100 Myr. 
  \end{itemize}
  \item Different volatile cases show different evolution of their atmospheric vertical extent because of the coupled effects of varying volatile solubility, mantle crystallization, and surface condensation.
  \item Magma ocean protoatmospheres show distinctive features in their spectra for different volatiles and at different times. These may guide future observations to discriminate between atmospheric primary composition and evolutionary stage.
  \item Spectra are further distinctive depending on interior parameterizations that alter petrological and energetic properties of magma ocean evolution. Therefore, astronomical observations of magma ocean planets may inform further understanding of the geophysical and geochemical evolution of planetary mantles and surfaces.
\end{itemize}

In conclusion, our work illustrates the importance of coupled models of rocky planetary mantles and their overlying atmospheres to reveal crucial feedback cycles that occur during early planetary evolution. Given the variety of ways to represent the physics and chemistry involved, there is a need for a range of models incorporating various aspects of the problem. We have introduced a novel model that adequately treats the primary energy transport mechanisms operating in a co-evolving planetary mantle and atmosphere. Future work will build on this modeling framework and pave the way to a more complete understanding of the environmental and surface conditions on young, rocky worlds. We anticipate that further developments---such as laboratory constraints on outgassing, atmospheric chemistry, and cloud feedbacks---will allow to simulate the extensive range of the rocky planet phase space. The insights gained through this venture will enhance our interpretation of the exoplanet census and further constrain the earliest physico-chemical conditions of our own world.

\acknowledgments

The authors thank P. A. Sossi and S. P. Quanz for comments on earlier drafts, A. G. J{\o}rgensen for providing additional line lists, L. Lichtenberg for tireless efforts in sampling petrological literature, and X. Tan, J. Wade, V. Parmentier, and R. J. Graham for discussions. This work was supported by grants from the Simons Foundation (SCOL award \#611576 to T.L.), the Swiss National Science Foundation (Early Postdoc.Mobility fellowship \#P2EZP2-178621 to T.L., Ambizione grant \#173992 to D.J.B.), and the European Research Council (Advanced grant
EXOCONDENSE \#740963 to R.T.P.). \textsc{data availability:} The simulation data and plotting scripts to reproduce the presented findings are available at \citet[osf.io/m4jh7]{dataA,dataB,dataC}. \textsc{software:} \textsc{spider} \citep{2018PEPI..274...49B}, \textsc{socrates} \citep{edwards1996socrates}, \textsc{numpy} \citep{Numpy2020,numpy_v1.19.1}, \textsc{scipy} \citep{2020SciPy-NMeth,pauli_virtanen_2020_3958354}, \textsc{pandas} \citep{pandas:2010,jeff_reback_2019_3509135}, \textsc{matplotlib} \citep{Hunter:2007,thomas_a_caswell_2019_3264781}, \textsc{seaborn} \citep{seaborn:2018}.

\clearpage

\appendix

\renewcommand{\thefigure}{A\arabic{figure}}
\renewcommand{\thetable}{A\arabic{table}}
\setcounter{figure}{0}
\setcounter{table}{0}

\section{Opacity data}

\begin{table}[htb]
\centering
\begin{tabular}{llrrl}
Molecule / system & Type & $\nu_{\mathrm{min}}$ (cm$^{-1}$) & $\nu_{\mathrm{max}}$ (cm$^{-1}$) & Reference\\ \hline
H$_2$O & lbl & 0 & 25711 & \citet{HITRAN2016} \\
H$_2$O & continuum & & & \citet{2012RSPTA.370.2520M} \\
CO$_2$ & lbl & 158 & 140757 & \citet{HITRAN2016} \\
CO$_2$--CO$_2$ & CIA & 1 & 2850 & \citet{1997Icar..129..172G} \\
H$_2$--H$_2$ & CIA & 20 & 20000 & \citet{borysow2001high,borysow2002collision} \\
CH$_4$ & lbl & 0 & 11502 & \citet{HITRAN2016} \\
CH$_4$--CH$_4$ & CIA & 1 & 990 & \citet{1987ApJ...318..940B} \\
N$_2$ & lbl & 12 & 9354 & \citet{HITRAN2016} \\
N$_2$--N$_2$ & CIA & 0 & 5000 & \citet{2015JChPh.142h4306K} \\
O$_2$ & lbl & 0 & 17272 & \citet{HITRAN2016} \\
O$_2$--O$_2$ & CIA & 1150 & 29800 & \citet{2004JMoSp.228..432B} \\
CO & lbl & 4 & 1447 & \citet{HITRAN2016} \\
\end{tabular}
\caption{\textsf{List of molecules and references for the opacity data used in this work. Abbreviations: lbl -- line-by-line, CIA -- collision-induced absorption.}}
\label{tab:A1}
\end{table}

\bibliography{references}{}
\bibliographystyle{aasjournal}

\end{document}